\documentclass{aa}
\usepackage{times}
\usepackage{txfonts}
\usepackage{graphicx}
\usepackage{xspace}
\usepackage{epsfig}
\usepackage{natbib}
\usepackage{rotating}
\usepackage{dcolumn}

\newcommand{\lglxlbol}{$\log{(L_{\rm x}/L_{\rm bol})}$}

\begin{document}

\title{X-ray emission of brown dwarfs: \\ Towards constraining the dependence on age, luminosity, and temperature}

\author{B. Stelzer\inst {1,2} \and G. Micela\inst {2} \and E. Flaccomio\inst {2} \and R. Neuh\"auser\inst {3} \and R. Jayawardhana\inst {4}}

\offprints{B. Stelzer}

\institute{Dipartimento di Scienze Fisiche ed Astronomiche, 
% INST 1
Universit\`a di Palermo, Piazza del Parlamento 1, I-90134 Palermo, Italy \email{B. Stelzer, stelzer@astropa.unipa.it}
\and 
% INST 2
INAF - Osservatorio Astronomico di Palermo,
  Piazza del Parlamento 1, I-90134 Palermo, Italy \and 
% INST 3
Astrophysikalisches Institut und Universit\"ats-Sternwarte Jena, Schillerg\"asschen 2-3, D-07745 Jena, Germany \and
% INST 4
Department of Astronomy \& Astrophysics, University of Toronto, Toronto, ON\,M5S\,2H8, Canada}

\titlerunning{X-ray emission from Brown Dwarfs}

\date{Received $<$date$>$ / Accepted $<$date$>$} 

\abstract
{}
{We observed brown dwarfs in different evolutionary stages 
with the Chandra X-ray Observatory with the aim to disentangle
the influence of different stellar parameters
on the X-ray emission of substellar objects. 
The ages of our three targets (HR\,7329\,B, Gl\,569\,Bab, and HD\,130948\,BC) 
are constrained by them being companions to main-sequence stars of known age. 
With both known age and effective temperature or bolometric luminosity, the mass can be 
derived from evolutionary models.
}
{Combining the new observations with previous studies presented in the literature
yields a brown dwarf sample that covers the age range from 
$\sim 1$\,Myr to $\sim 1$\,Gyr. %, and reaches well down into the cooling phase of brown dwarfs.
Since the atmospheric temperature of brown dwarfs is approximately constant at young ages,  
a sample with a large age spread is essential for investigating the possible influence 
of effective temperature on X-ray activity. 
}
{Two out of three brown dwarfs are detected with {\em Chandra}, with variable lightcurves and comparatively
soft spectra. 
Combining our results with published data %on X-ray emission from brown dwarfs 
allows us to consider a subsample of high-mass brown dwarfs (with $0.05-0.07\,M_\odot$), 
thus eliminating mass from the list of free parameters. 
We find evidence that X-ray luminosity declines with decreasing bolometric luminosity steeper than 
expected from the canonical relation for late-type stars ($L_{\rm x}/L_{\rm bol} = 10^{-3...-5}$). 
Effective temperature is identified as a likely parameter responsible for the additional decline of X-ray
activity in the more evolved (and therefore cooler) brown dwarfs of the `high-mass' sample. 
In another subsample of brown dwarfs characterized by similar effective temperature,  
the X-ray luminosity scales with the bolometric luminosity without indications for a deviation from 
the canonical range of $10^{-3...-5}$ observed for late-type stars.} 
{Our findings support the idea that effective temperature plays a critical role for the X-ray 
activity in brown dwarfs.
This underlines an earlier suggestion based on observations of chromospheric H$\alpha$
emission in ultracool dwarfs that the low ionization fraction in the cool brown dwarf atmospheres may 
suppress magnetic activity.} 


\maketitle

\section{Introduction}\label{sect:intro}

In the last few years a large number of very low-mass (VLM) dwarf stars and brown dwarfs (BD)
have been discovered in the solar neighborhood by sky surveys such as
2\,MASS, DENIS, or SDSS. Spectroscopic follow-up of these objects has enabled the first
systematic investigation of chromospheric H$\alpha$ activity in stars of the very smallest
masses and even BDs \citep{Gizis00.1, Mohanty03.1}. One of the most
important results of these studies is that H$\alpha$ emission seems to reach a maximum at
spectral type M7, and fade off for the latest M dwarfs suggesting a change in
the efficiency of the mechanism that drives magnetic activity.
 
X-ray emission is a complementary activity indicator for late-type stars, probing
the interplay between magnetic field and matter in the outermost and hottest part of the
atmosphere, the corona.
\cite{Huensch99.1} have presented a study of X-ray emission on field M dwarfs
based on the {\em ROSAT} All-Sky Survey providing detections for spectral type
earlier than M7. But beyond spectral type M6 the X-ray regime is widely unexplored.
Only a handful of late M dwarfs have been detected with the {\em EXOSAT}
and/or {\em ROSAT} missions.
Except LHS\,2065, for which \cite{Schmitt02.1}
detected quiescent emission with the {\em ROSAT} HRI, all these objects
were only detected during a flare
\citep{Giampapa96.1,Fleming00.1}.
The coolest field object detected in X-rays so far is
LP\,944-20, an intermediate age M9.5 BD \citep[$t \sim 500$\,Myrs;][]{Tinney98.1}. %, which
While {\em ROSAT} had provided a comparatively high upper limit
\citep{Neuhaeuser99.1}, {\em Chandra} has
seen LP\,944-20 during a flare \citep{Rutledge00.1}, and {\em XMM-Newton} has
provided the lowest upper limit for the quiescent flux of any field dwarf so far:
$L_{\rm x} < 3.1 \times 10^{23}\,{\rm erg/s}$ \citep{Martin02.1}.
Deep X-ray studies have been attempted for only few L dwarfs: 
Denis\,J1228-1547 (an L5 binary) remained undetected in an {\em XMM-Newton}
observation, which was effectively quite shallow due to high background resulting from 
solar flares \citep{Stelzer02.1}. \citet{Berger05.1} presented
{\em Chandra} observations of two nearby L dwarfs, 
an upper limit of $L_{\rm x} < 6.6 \times 10^{24}\,{\rm erg/s}$ for 2MASS\,J1507-1627 (SpT L3.5) 
and of $L_{\rm x} < 9.3 \times 10^{24}\,{\rm erg/s}$ for 2MASS\,J0036+1821 (SpT L5).
The latter one was observed simultaneously in various wavebands including radio, optical
and X-rays. Finally, the T dwarf binary $\epsilon$\,Ind\,Bab was observed but not
detected with {\em Chandra}, yielding an upper limit of 
$L_{\rm x} < 3.2 \times 10^{23}\,{\rm erg/s}$ \citep{Audard05.1}.
                                                                                
While the observational material on X-ray activity
of ultra-cool field dwarf stars and BDs is scarse,
X-ray emission has lately been revealed from a substantial number of $\sim 1-5$\,Myr-old
VLM pre-MS stars and BDs in various star forming regions \citep[e.g.][]{Neuhaeuser98.1, Preibisch02.1,
Mokler02.1, Feigelson02.1, Getman02.1, Stelzer04.1, Preibisch05.1},
for one BD in the $\sim 10$\,Myr-old TW\,Hya association \citep{Tsuboi03.1},
one $\sim 5$\,Myr old member of Upper Sco \citep{Bouy04.1}, and one BD in the Pleiades \citep{Briggs04.1}.

BDs are not able to fuse normal hydrogen and subsequently must cool down and become fainter
as they age. Therefore, their detection at young ages but non-detection in more
evolved stages suggests a connection between activity and effective temperature
or bolometric luminosity.
Based on the absence or low level of H$\alpha$ emission seen in most
field L dwarfs \citet{Mohanty02.1} have argued that the
chromospheric emission may shut off below a critical
temperature because the atmosphere becomes too neutral to provide substantial
coupling between matter and magnetic field. But due to a lack of systematic
observations this hypothesis has not been tested so far.
On the other hand, the X-ray luminosity of late-type stars in the solar neighborhood 
with spectral type earlier than $\sim$\,M6 is seen to decrease along the spectral type sequence,
while $L_{\rm x}/L_{\rm bol}$ remains approximately
constant at $\sim -4$ with a spread of $\sim 2$\,dex for each spectral type
subclass; see e.g. Fig.~4 in \citet{Stelzer04.2}. 
If the $L_{\rm x} - L_{\rm bol}-$relation of late-type stars
proceeds into the cool end of the MS and beyond the substellar limit
the non-detection of most VLM field stars and BDs with the  previously available 
low-sensitivity X-ray instruments is no surprise.
However, as we will show in this paper {\em Chandra} is capable of providing meaningful
constraints for the X-ray emission of these faint objects.

\section{Sample}\label{sect:sample}

We have embarked on a systematic {\em Chandra} study of VLM stars and BDs
with the aim to 
constrain the influence of stellar parameters on their X-ray emission. 
One of the pre-requisites for our project is to know the age and effective temperature 
and/or bolometric luminosity of the BDs.
The mass can then be deduced from evolutionary models.
Both $T_{\rm eff}$ and $L_{\rm bol}$ decrease rapidly with age. Therefore, is it
important to examine {\em evolved} BDs, and compare their X-ray properties to those
of {\em young} BDs in star forming regions. To this end, we have compiled from the literature
X-ray luminosities and stellar parameters for BDs in various evolutionary stages, and combine
these data with our new {\em Chandra} observations.   

\begin{figure}
\begin{center}
\parbox{8cm}{\resizebox{8cm}{!}{\includegraphics{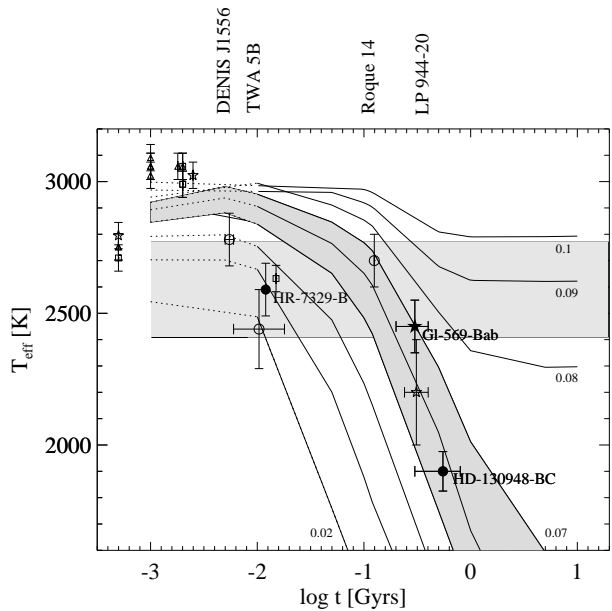}}}
\caption{BDs and BD candidates with known age 
on the model tracks from \protect\citet{Baraffe98.1} and \protect\citet{Chabrier00.1} 
in steps of $0.01\,M_\odot$.
The $0.08\,M_\odot$ track marks the separation from stars to BDs.
Only objects that were detected with {\em XMM-Newton} or {\em Chandra} are shown. 
These include:
{\em squares} - BDs in IC\,348; 
{\em triangles} - BDs in Cha\,I; 
{\em open circles} - further BDs compiled from the literature, see labels on top of the figure; 
{\em filled symbols} - BDs observed in the {\em Chandra} program described in this paper. 
Objects on which an X-ray flare was detected are marked by a star symbol. 
The grey-shades indicate distinct mass and temperature ranges we study in this paper.} 
\label{fig:teff_age}
\end{center}
\end{figure}

Fig.~\ref{fig:teff_age} shows the socalled cooling curves (effective temperature
versus age for objects of given mass) according to the model calculations
by \citet{Baraffe98.1} and \citet{Chabrier00.1}.  
The data points represented by open plotting symbols refer to BDs from {\em XMM-Newton} or {\em Chandra}
observations compiled from the literature. For clarity we show only BDs that were {\em detected} 
with {\em XMM-Newton} or {\em Chandra}. Non-detections are not displayed in Fig.~\ref{fig:teff_age}, 
but will be included in the discussion later on. 
Filled plotting symbols denote targets from our new {\em Chandra} observations. 
As summarized in Sect.~\ref{sect:intro}, previous X-ray satellites did not yield meaningful
constraints on BD X-ray emission, with the exception of some very young BDs, 
and their results will not be repeated here. 

As is clear from Fig.~\ref{fig:teff_age} the cooling curves are poorly populated with 
X-ray data, especially at evolved ages. 
Several of the X-ray observed BDs mentioned in Sect.~\ref{sect:intro} do not show up in 
Fig.~\ref{fig:teff_age} because their age is not known.  
This is because the age determination in the coolest part of
the HR diagram is subject to considerable uncertainty for two reasons: 
(i) For BDs there is always a degeneracy in the mass-age plot because
they do not reach the MS. (ii) Theoretical evolutionary models for VLM
stars and BDs start with assumed initial conditions,
so that the results are quite uncertain for the first $10$ to $100$\,Myrs,
and they have not been tested nor calibrated for VLM objects.
%
% OUTPUT FROM     plot_teff_age.pro,  plot_lbol_age.pro
%

To circumvent the problem of the age determination for the BDs 
we have selected our {\em Chandra} sample among substellar companions to MS stars.
Ages of MS stars can be estimated by a number of methods, e.g. space motion, metallicity,
rotation, and activity \citep[e.g.][]{Lachaume99.1}, and companions
are expected to be coeval.
Within our program {\em Chandra} has observed three BD companions 
that span a significant range of ages ($\sim 12 - 550$\,Myrs),
$T_{\rm eff}$ ($2600 - 1900$\,K), $\log{L_{\rm bol}}$ ($29.7 - 31.0$\,erg/s), 
and masses ($0.02-0.07\,M_\odot$); 
filled circles in Fig.~\ref{fig:teff_age}. %, 
The youngest of them, HR\,7329\,B, %at an age of $\sim 20$\,Myr
connects to the sequence of BDs in young stellar associations and star forming regions. 
The other two targets (HD\,130948\,BC and Gl\,569\,Bab)
have similar age, % and mass, 
but $T_{\rm eff}$ differs by $\sim 500$\,K (corresponding to $\sim 5$ spectral subclasses) 
and $L_{\rm bol}$ differs by a factor of $\sim 20$, 
as a result of the strong dependence of these parameters on mass for evolved BDs. 
The {\em Chandra} observation of Gl\,569\,Bab has already been discussed in detail by \citet{Stelzer04.2}.
We include here the previous results, in addition to some further X-ray parameters not given
in \citet{Stelzer04.2}. 
The stellar parameters and X-ray properties of all BD targets are summarized in Table~\ref{tab:sample}. 
\begin{table*}
\begin{center}
\caption{Sample of BD companions to MS stars observed with {\em Chandra} ACIS-S. 
The left hand side of the table represents a summary of stellar parameters. The distances
are from {\em Hipparcos}, references for the other parameters are given in brackets. 
On the right hand side we list X-ray properties derived from the {\em Chandra} data: 
X-ray luminosity, ratio of X-ray to bolometric luminosity attributing all X-ray emission
to either of the components for unresolved binaries, and median photon energy. The labels
`A', `Q', and `F' denote average, quiescent, and flare luminosities.} 
\label{tab:sample}
\small
\begin{tabular}{llrrrrrrrl|crrr} \hline
\multicolumn{10}{c}{} & \multicolumn{2}{|c}{[0.5-8\,keV]} \\
Name            & \multicolumn{1}{c}{Sep}              & \multicolumn{1}{c}{P.A.}      & Dist                     &  SpT  & \multicolumn{1}{c}{$M$}               & \multicolumn{1}{c}{$T_{\rm eff}$} & \multicolumn{1}{c}{Age} & \multicolumn{1}{c}{$\log{\frac{L_{\rm bol}}{L_\odot}}$} & Ref. & & $\log{L_{\rm x}}$ & $\log{\frac{L_{\rm x}}{L_{\rm bol}}}$ & $\langle E_{\rm ph} \rangle$ \\
                & \multicolumn{1}{c}{[$^{\prime\prime}$]} & \multicolumn{1}{c}{[$^\circ$]} & \multicolumn{1}{c}{[pc]} &       & \multicolumn{1}{c}{[${\rm M_\odot}$]} & [K]                               & [Gyrs]                  &                                                         &      & & [erg/s]           &                                       &                              \\ \hline
HR\,7329\,B         &  $4.097$ & $166.9$ & $47.7$  & M7.5  & $0.020-0.045$ & $2600$ & $0.012$     & $-2.59$ & (1,2)  & A & $27.2$   & $-3.8$   & $0.87$ \\
Gl\,569\,Bab        &  $4.89$  &  $30.0$ &  $9.8$  &       &               &        & $0.2-0.4$   &         & (3)    & A & $26.8$   &          & $0.85$ \\
                    &          &         &         &       &               &        &             &         &        & Q & $25.8$   &          & $0.67$ \\
                    &          &         &         &       &               &        &             &         &        & F & $27.5$   &          & $0.87$ \\
\hspace*{1cm} Ba    &          &         &         & M8.5  & $0.055-0.087$ & $2450$ &             & $-3.35$ & (4)    & A &          & $-3.2$   &        \\
\hspace*{1cm} Bb    &          &         &         & M9    & $0.034-0.070$ & $2450$ &             & $-3.58$ & (4)    & A &          & $-3.4$   &        \\
HD\,130948\,BC       &  $2.64$  & $104.5$ & $17.9$  &       &               &        & $0.3-0.8$  &         & (5)    & A & $<25.6$  &          & $-$    \\
\hspace*{1cm} B$^*$  &          &         &         & L4    & $0.040-0.065$ & $1900$ &            & $-3.81$ & (5,6)  & A &          & $<-4.2$  &        \\
\hspace*{1cm} C$^*$  &          &         &         & L4    & $0.040-0.065$ & $1900$ &            & $-3.93$ & (5,6)  & A &         & $<-4.1$  &        \\
\hline
\multicolumn{14}{l}{References: (1) - \citet{Guenther01.1}, (2) - \citet{Lowrance00.1}, (3) - \citet{Lane01.1}, (4) - \citet{Zapatero04.1},} \\
\multicolumn{14}{l}{(5) - \citet{Potter02.1}, (6) - \citet{Goto02.1};} \\
\multicolumn{14}{l}{$^*$ Bolometric luminosity estimated from the $K$ band magnitude and the bolometric correction for the $K$ band given by \cite{Leggett01.1}.} \\
\end{tabular}
\end{center}
\end{table*}

We complement this data set by BDs presented in the {\em XMM-Newton} and {\em Chandra} literature, 
for which the age is known. The X-ray literature comprises a number of nearby low-mass star forming regions. 
We include in our study 
only IC\,348 and Cha\,I. The stellar populations of these two regions were shown to be   
compatible with the \cite{Baraffe98.1} and \cite{Chabrier00.1} models.
For somewhat younger regions, such as $\rho$\,Oph and ONC, many objects lie above the youngest
isochrone in these models ($1$\,Myr), such that we can not place them in Fig.~\ref{fig:teff_age}. 
Furthermore, the stellar parameters for BDs in IC\,348 and Cha\,I presented in the literature, 
(spectral type, effective temperature and bolometric luminosity) 
were derived in the same manner for both regions, thus yielding a homogeneous data set. 
Specifically, the spectral types were determined by comparing the optical low-resolution
spectra to those of standard dwarfs and giants. And the temperature scale used for the conversion of spectral 
types to $T_{\rm eff}$ was devised by \citet{Luhman03.1} for use
with the \citet{Baraffe98.1} and \citet{Chabrier00.1} models. 

Our X-ray selected sample of IC\,348 is from \citet{Preibisch02.1} and that of Cha\,I from
\citet{Stelzer04.1}. 
We consider objects with spectral type M7 and later as bona-fide BDs, and those with
spectral type M5 to M7 as BD candidates. For simplicity we refer to both of them as `BDs' throughout
this paper. 
For IC\,348 the stellar parameters 
are taken from \citet{Luhman99.1} and \citet{Luhman03.1}. In the latter paper Luhman established
that four of the BD candidates in IC\,348 identified by \citet{Najita00.1} and included in the
X-ray study of \cite{Preibisch02.1} are possibly field stars (N021-05, N024-02, N071-02, N075-01). 
Furthermore, N013-05 is classified as an M3 star by \citet{Luhman03.1}, and therefore not a BD. We exclude
these five objects from our study.  
For Cha\,I we use the stellar parameters from the compilation by \citet{Luhman04.1}.  
In order to put the young BDs in Fig.~\ref{fig:teff_age} we estimated ages on an 
individual basis from the position of each object on the \citet{Baraffe98.1} and \citet{Chabrier00.1} 
isochrones in the HR diagram. BDs above the youngest isochrone are arbitrarily placed at an age of 
$0.5$\,Myr in Fig.~\ref{fig:teff_age},  
because for such young objects the evolutionary calculations are sensitive to the (unknown)
initial conditions and therefore unconstrained \citep{Baraffe02.1}. 
Actually, the models are uncertain for ages up to a few Myr. To indicate that the tracks
for young ages must be used with caution, they are drawn with dotted lines in Fig.~\ref{fig:teff_age}.
We stress that the results derived in our study do not depend sensitively on these uncertainties;
see discussion in Sect.~\ref{subsect:disc_xrayprop_bds}. 

Besides these young BDs in star forming regions {\em XMM-Newton} or {\em Chandra} data have been 
published for the following BDs with reliable age estimates from their membership in associations
or clusters: 
DENIS\,J1556 in Upper Sco ($5-6$\,Myr), TWA\,5B and 2M\,1207-39 in TW\,Hya ($\sim 10$\,Myr),  
and five BDs in the Pleiades ($\sim 125$\,Myr); see Sect.~\ref{sect:intro} for references to the X-ray data. 
In the solar neighborhood the number of BDs and VLM stars with known age is very limited,
due to the problems outlined above. 
X-ray data have been presented for LP\,944-20, APMPM\,J2354, and $\epsilon$\,Ind\,Bab. 
For LP\,944-20 we adopt an age of $320 \pm 80$\,Myr, derived from its association with 
the Castor moving group \citep{Ribas03.1}. Note that this is lower than the previous age estimate 
based on its lithium equivalent width \citep{Tinney98.1}.
APMPM\,J2354 is a proper motion companion to a binary composed of a M4 dwarf star 
and a white dwarf \citep{Scholz04.1}.  
The age of the system is $\sim 1.8$\,Gyr determined from the cooling history of 
the white dwarf \citep{Silvestri01.1}. 
APMPM\,J2354 is the only star in our sample, and it is included in this study 
in order to examine the X-ray properties
across the substellar boundary. The T dwarf binary $\epsilon$\,Ind\,Bab has an age of
$0.2-2$\,Gyr estimated from the rotation properties of the primary $\epsilon$\,Ind\,A \citep{Lachaume99.1}. 

For the objects introduced in the previous paragraph the bolometric luminosity and the spectral
type are compiled from the literature. We have derived their effective temperature 
from the spectral type using the scale given by \citet{Mohanty03.1}. We caution that the
temperatures of the younger BDs ($\leq 100$\,Myr) with lower gravity 
may be somewhat higher than inferred with this scale which was devised for dwarfs. But within the estimated
uncertainties this does not affect our study. The values for $T_{\rm eff}$ obtained this way are
compatible with the temperatures predicted by the \citet{Baraffe98.1} and \citet{Chabrier00.1}
models, except for LP\,944-20, where there is a discrepancy of $400$\,K; see also discussion 
in \citep{Ribas03.1}. For this BD we assume a temperature of $2200 \pm 200$\,K, 
intermediate between the empirical scale and the evolutionary calculations.

\section{Observations and Data Analysis}\label{sect:obs}

\begin{table}
\begin{center}
\caption{Observing log.}
\label{tab:obslog}
\begin{tabular}{lcrr}\hline
Target         & Obs-ID & \multicolumn{1}{c}{UT Start [hh:mm:ss]} & Expo [ks] \\ \hline
Gl\,569\,Bab   & 4470 & 2004-06-18~~05:59:44 & 25.2 \\
HD\,130948\,BC & 4471 & 2004-11-24~~07:45:33 & 42.5 \\
               & 6166 & 2004-11-27~~09:22:07 &  7.3 \\
HR7329\,B      & 4472 & 2004-08-06~~12:44:55 & 47.0 \\ \hline
\end{tabular}
\end{center}
\end{table}
The observing logs for our {\em Chandra} observations are summarized in Table~\ref{tab:obslog}. 
All three targets were observed with the Advanced CCD Imaging Spectrometer (ACIS),
using the S3 chip in imaging mode. 
The data reduction for HR\,7329 and HD\,130948 was carried out using the CIAO software
package\footnote{CIAO is made available by the CXC and can be downloaded
from \\ http://cxc.harvard.edu/ciao/download-ciao-reg.html} version 3.2
in combination with the calibration database (CALDB) version 3.0.0, 
in an analogous way to the analysis of the data for Gl\,569 
which was already presented by \citet{Stelzer04.2}. 

We started the analysis with the level\,1 events file provided by the
pipeline processing at the {\em Chandra} X-ray Center (CXC).
In the process of converting the level\,1 events file to a level\,2 events file
for each of the observations the following steps were performed:
Charge transfer inefficiency (CTI) correction,
removal of the pixel randomization,
filtering for event grades (retaining the standard {\em ASCA} grades $0$, $2$, $3$,
$4$, and $6$),
and application of the standard good time interval (GTI) file.
The events file was also checked for any systematic aspect offsets using
CIAO software, but none were present.
For HD\,130948 the level\,2 events files of the two observations were merged for
the purpose of source detection. 
Source detection was carried out with the {\it wavdetect} algorithm
on an image of $50 \times 50$\,pixels length ($1$\,pixel = $0.492^{\prime\prime}$)
centered on the computed position of the primary
using wavelet scales between $1$ and $8$ in steps of $\sqrt{2}$.

As a consequence of the small binary separations of our targets ($\sim 2.5-5^{\prime\prime}$), 
the cross-correlation between X-ray and optical/IR positions requires precise astrometry. 
We translated the {\em Hipparcos} position of the primary star to the time of the 
{\em Chandra} observation using its proper motion \citep{Perryman97.1}. 
The position of the companion relative to the primary was then 
obtained using the separation and P.A. given in the literature (see 
Table~\ref{tab:sample}).

\begin{figure*}[t]
\begin{center}
\parbox{17cm}{
\parbox{5.5cm}{\resizebox{5.5cm}{!}{\includegraphics{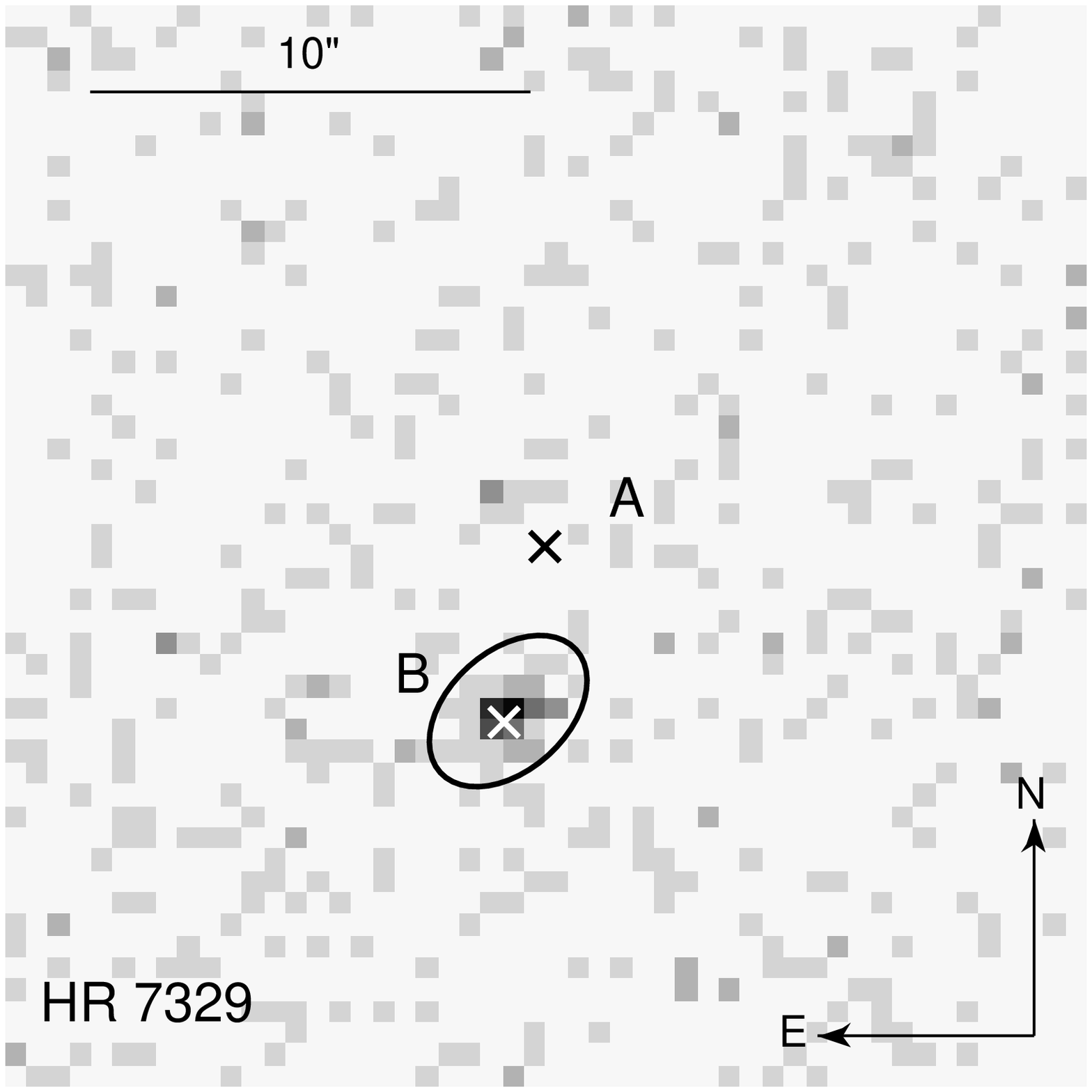}}}
\parbox{5.5cm}{\resizebox{5.5cm}{!}{\includegraphics{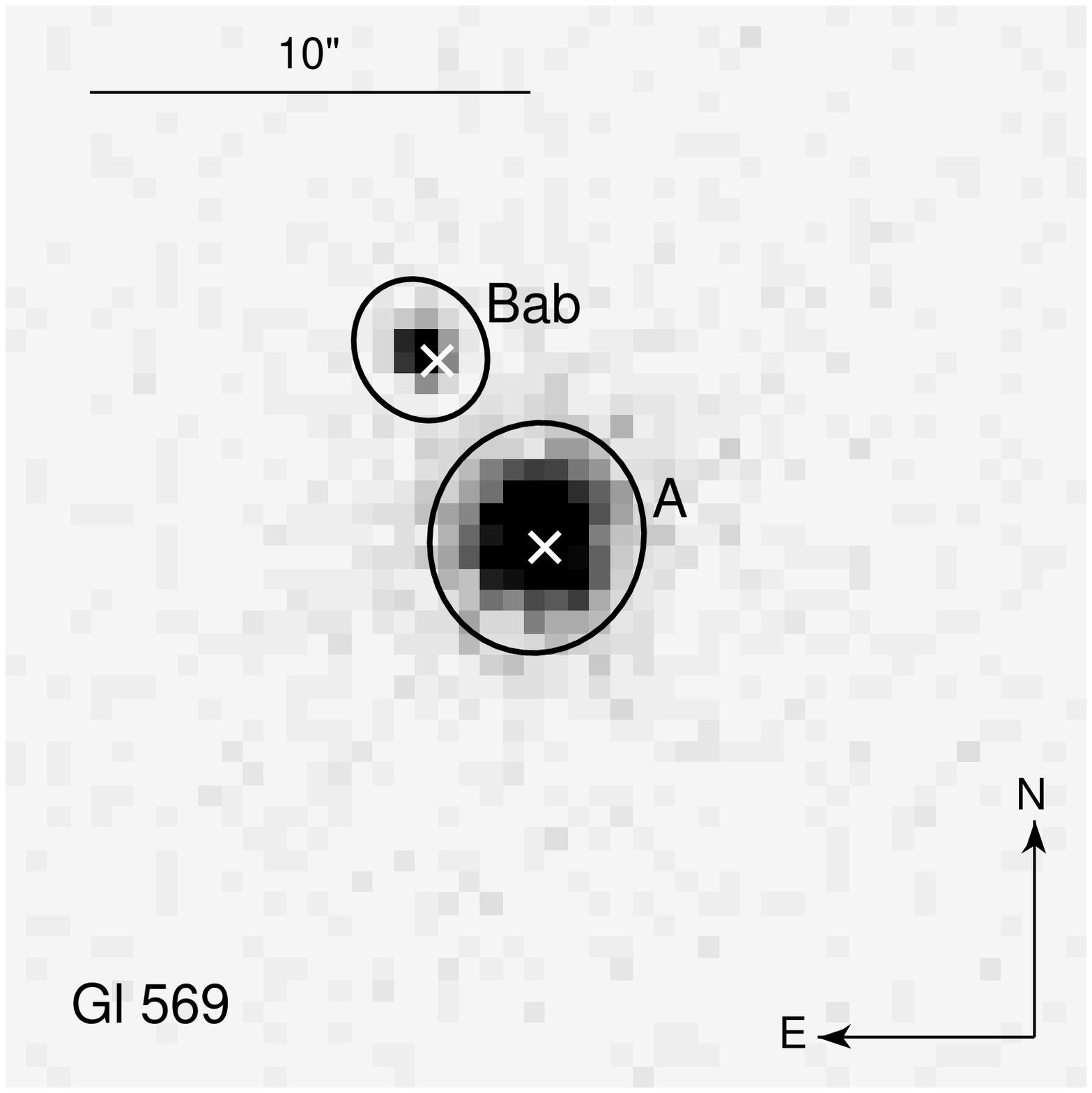}}}
\parbox{5.5cm}{\resizebox{5.5cm}{!}{\includegraphics{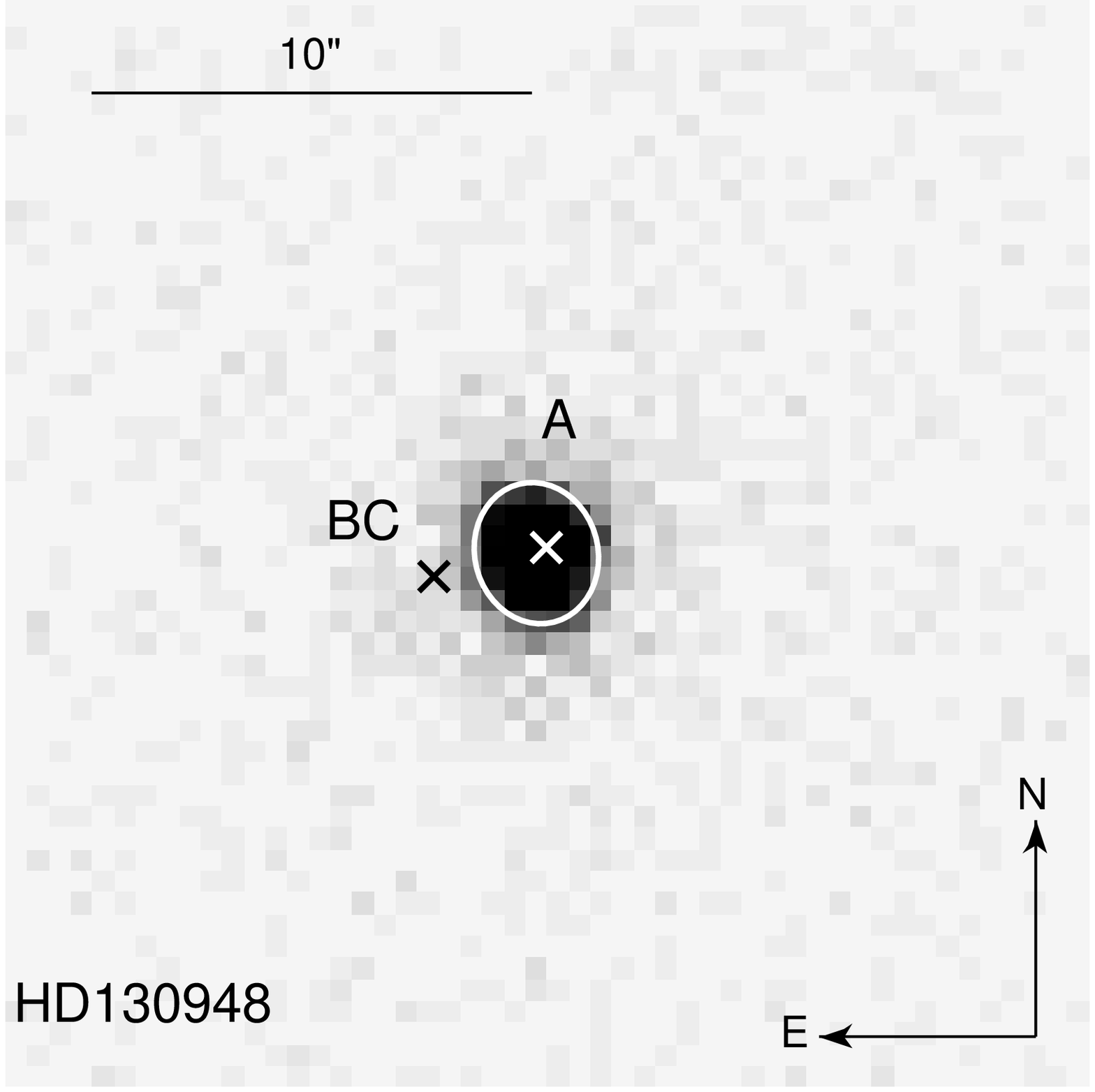}}}
}
\caption{X-ray images of stellar systems with BD secondary observed with {\em Chandra} ACIS-S: 
{\em left} - HR\,7329 ($12$\,Myr), {\em middle} - Gl\,659 ($300$\,Myr), {\em right} - HD\,130948 ($550$\,Myr). The primaries are labeled `A'.}
\label{fig:acis_images}
\end{center}
\end{figure*}

In Fig.~\ref{fig:acis_images} we show the corresponding {\em Chandra} ACIS images
for all three targets. X-ray positions are indicated by the $3\,\sigma$ source ellipse of {\em wavdetect}, 
and optical positions are marked by crosses. We label the primary with `A', and the companions 
with `B'. (In two of the targets the companion is itself a binary, composed of two BDs, but with
separations below {\em Chandra}'s resolution limit, cf. Table~\ref{tab:sample}.) 
%
% OUTPUT FROM    ds9
%

We carefully selected individual photon extraction areas for each component of the three targets. 
For the BDs circular regions were chosen with radius determined by the separation to the primary 
and by the X-ray brightness of the primary. 
Two of the primaries (Gl\,569\,A and HD\,130948\,A) are very strong X-ray emitters, and pile-up
is not negligible. For these two sources we use annular photon extraction regions, where the inner
radius is chosen such as to minimize the loss of statistics while removing all pile-up (see
Sect.~\ref{subsect:hd130948} and \cite{Stelzer04.2} for details). 
The third primary (HR\,7329\,A) is not detected. 
The background was selected for each target from carefully defined regions near the X-ray positions.
In particular, in the case of HD\,130948\,BC, which is located in the wings of the PSF of its
primary, we extracted the background from an area on the opposite side of and with the same separation
from the primary.
For undetected objects upper limits to the source counts were computed 
following the prescription for Poisson distributed data given by \cite{Kraft91.1}.

For objects that are undetected or too faint for spectral analysis X-ray fluxes are estimated with 
PIMMS\footnote{PIMMS is available at http://asc.harvard.edu/toolkit/pimms.jsp} 
from the measured count rates assuming an unabsorbed isothermal Raymond-Smith plasma \citep{Raymond77.1}.
The counts-to-energy conversion factor is independent of the temperature
in the range of $0.3-1$\,keV which is typical for coronal plasmas of VLM stars and BDs 
\citep[see also discussion by][]{Audard05.1}. 
Finally, fluxes are converted to luminosities using the {\em Hipparcos} distances of the primaries. 
The X-ray luminosities and X-ray spectral parameters are listed in Table~\ref{tab:sample} for the BDs and in 
Table~\ref{tab:sample_prim} for the primaries. For lack of sufficient statistics we provide the
median photon energies for the BDs rather than a coronal temperature derived from spectral modeling.   
Next we discuss the results on an individual basis. 

\begin{table}
\begin{center}
\caption{X-ray parameters of primaries in observed systems.}
\label{tab:sample_prim}\small
\begin{tabular}{llr|rrr} \hline
\multicolumn{3}{c}{} & \multicolumn{2}{|c}{[0.5-8\,keV]} \\
Name            &  SpT  & \multicolumn{1}{c|}{$\log{\frac{L_{\rm bol}}{L_\odot}}$} & $\log{L_{\rm x}}$ & $\log{\frac{L_{\rm x}}{L_{\rm bol}}}$ & $\langle E_{\rm ph} \rangle$ \\
                &       &                                                         & [erg/s]           &                                       &                              \\ \hline
HR\,7329\,A      & A0   & $+1.37$ & $<26.1$  & $<-8.8$ & $-$ \\
Gl\,569\,A~Quies & dM2e & $-1.46$ & $ 28.3$  & $-3.8$  & $0.81$  \\
Gl\,569\,A~Flare & dM2e & $-1.46$ & $ 28.7$  & $-3.4$  & $0.90$  \\
HD\,130948\,A    & G2   & $+0.13$ & $ 28.8$  & $-4.9$  & $0.81$  \\
\hline
\end{tabular}
\end{center}
\end{table}

\section{Results}\label{sect:results}

\subsection{HR\,7329}\label{subsect:hr7329}

The Vega-like A0 star HR\,7329 %( + HD\,181296; HIP\,95261; $\eta$\,Tel) 
was first considered to be a member of the Tucanae association \citep{Zuckerman00.1},
but later shown to belong to the `$\beta$\,Pic moving group' \citep{Zuckerman01.1}.
This implies that its age has been revised downward (to $12$\,Myr instead of $\sim 40$\,Myr). 
 
HR\,7329\,B, at a separation of $4.2^{\prime\prime}$ from the A-type star,  
was discovered by \citet{Lowrance00.1} and a spectral type M7.5\,V was assigned, 
corresponding to substellar mass under the assumption that it forms a bound pair
with HR\,7329\,A. It was indeed established by \citet{Guenther01.1} that both objects  
share common proper motion. 

\begin{figure}[t]
\begin{center}
\resizebox{8.5cm}{!}{\includegraphics{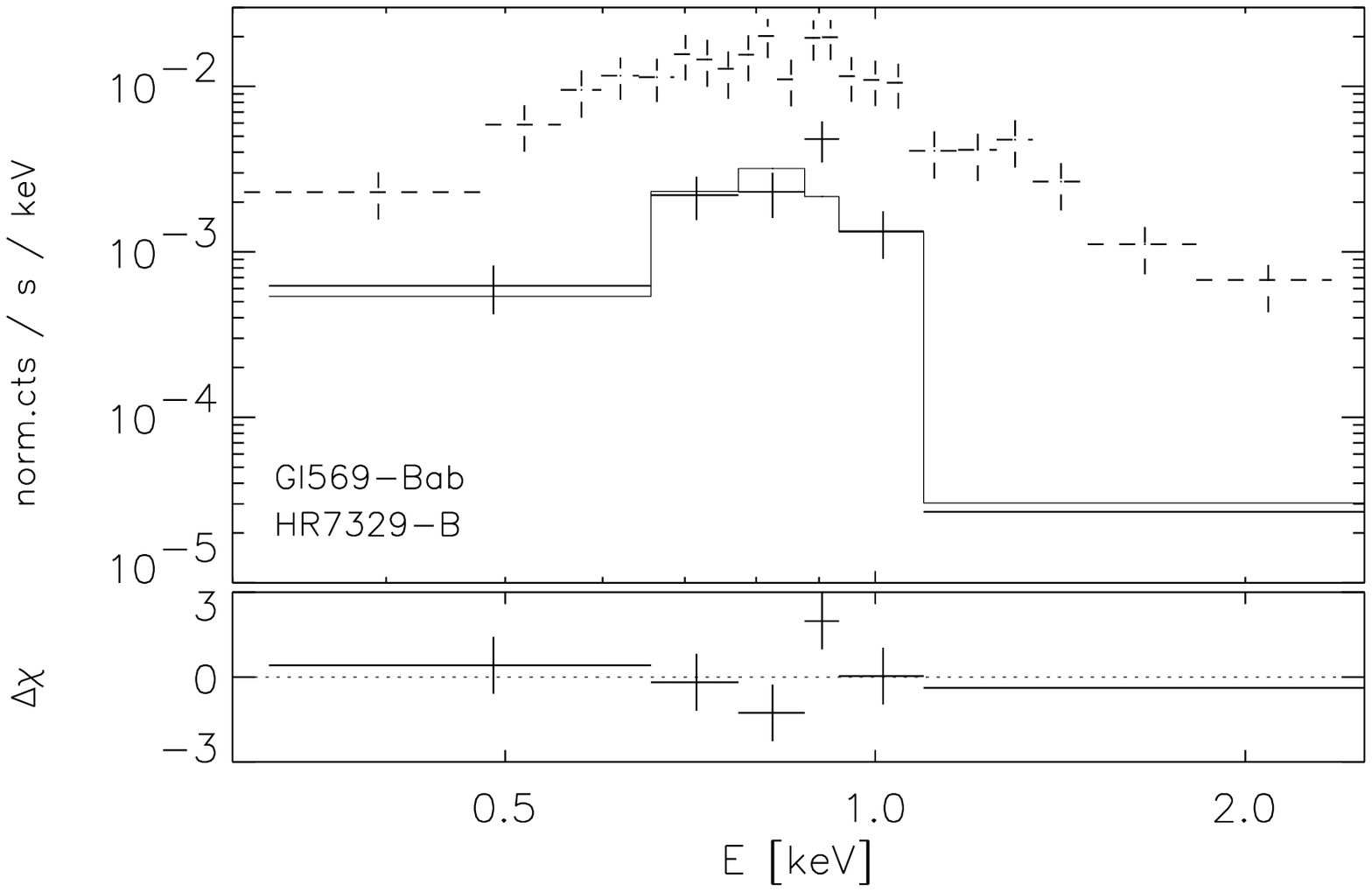}}
\resizebox{8.5cm}{!}{\includegraphics{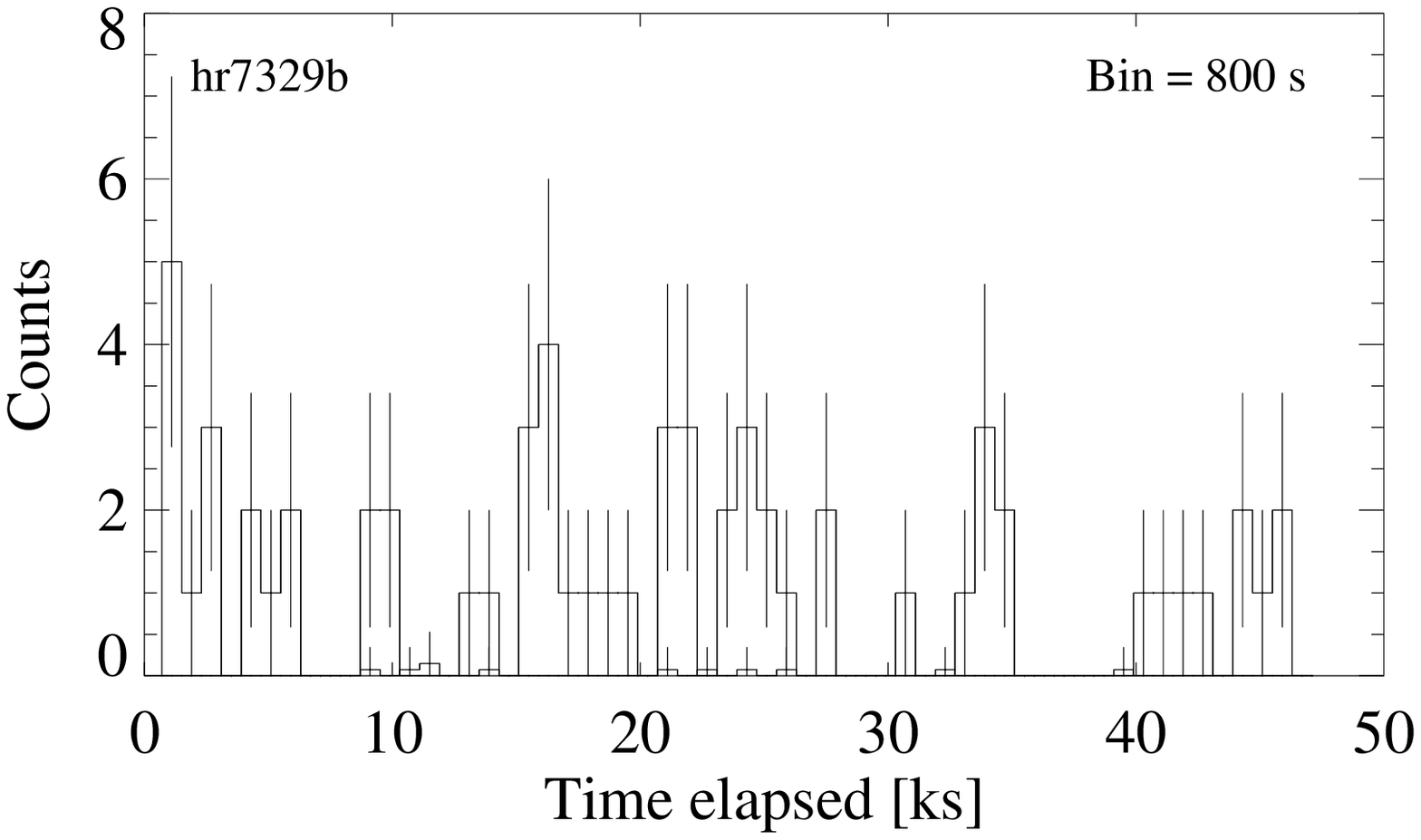}}
\caption{Top: {\em Chandra} spectrum of HR\,7329\,B, best fit 1-T APEC model, and residuals. 
The dashed lines represent the spectrum of Gl\,569\,Bab \protect\citep[see][]{Stelzer04.2}, 
shown for comparison. Bottom: {\em Chandra} lightcurve of HR\,7329\,B. The background, 
also shown and scaled to the source extraction area, is negligible.}
\label{fig:spec_lc_hr7329}
\end{center}
\end{figure}

In the {\em Chandra} image there is only one X-ray source at the position of the
binary, coinciding with the BD. 
The total number of source photons collected from HR\,7329\,B is $66.5$\,counts, after subtracting
the background, resulting in a time-averaged luminosity of $\log{L_{\rm x}} = 27.2$\,erg/s. 
% THIS lgLx IS FOUND BOTH WITH PIMMS AND FROM SPECTRAL MODELLING IN THE 0.5-8 KEV BAND.
The $\log{(L_{\rm x}/L_{\rm bol})}$ ratio is $-3.8$, within the typical range of late-type stars. 
The spectral shape is not well constrained due to the poor statistics.
In the top panel of Fig.~\ref{fig:spec_lc_hr7329} we show the X-ray spectrum of HR\,7329\,B 
overlaid by the best fit of a one-temperature ($1$-T) 
APEC\footnote{For spectral fitting we use the Astrophysical Plasma Emission Code within the
XSPEC environment version 11.3.0; see \cite{Smith01.1} for a description of APEC.} 
model with subsolar abundances ($Z = 0.3\,Z_{\odot}$). No photo-absorption term is required to fit
the spectrum. The temperature of the best fit model is $0.49$\,keV ($ = 5.7$\,MK).
%
% OUTPUT FROM    /home/stelzer/data/IDL/allg/plot_acis.pro   (prepare_specplot.pro)
%
%%
%% OUTPUT FROM   /home/stelzer/data/publ/bd_comps/ps/lc.pro 
%%                  with parameters: ipath='/um1/stelzer/chandra/bd_comps/200241_cd/data/'
%%                                   whichstar='hr7329b'
%%                                   whichcomp='B'
%%                                   extrarea='circ'
%%                                   xmin=0.0 & xmax=50.0 & ymax=8.0
%%
%

The lightcurve of HR\,7329\,B (bottom panel of Fig.~\ref{fig:spec_lc_hr7329}) 
is suggestive of variability. %, 
A KS-test applied to the photon arrival times yields a probability of $97$\,\% for variability.
Furthermore, we used the MLB algorithm, a method to determine periods of constant signal from a list of photon
arrival times, based on a maximum likelihood algorithm under the assumption of Poisson statistics;
see \citet{Wolk05.1} for a description of this technique. However, no change point is found at
significance $\geq 95$\,\%, i.e. the whole lightcurve can be described by constant count rate.   

The A0 star HR\,7329\,A is X-ray dark down to the detection limit, with an upper limit of 
$\log{L_{\rm x}} < 26.1$\,erg/s, corresponding to \lglxlbol$< -8.8$. 
% lglbol for HR7329A is from Stelzer & Neuhaeuser 2000, AA361
% where it was computed from the V-mag with the Schmidt-Kaler82 bolometric correction.
This is (one of the)
most sensitive upper limits for the X-ray emission from an A-star. It is by three (!) orders 
of magnitude more sensitive than the upper limit derived from the {\em ROSAT} All-Sky Survey
for the combined system HR\,7329\,AB \citep{Stelzer00.1}.

\subsection{HD\,130948}\label{subsect:hd130948}

HD\,130948\,A is a young solar analog (spectral type G2\,V). 
A comparatively young age ($0.3-0.8$\,Gyr) is indicated by 
high lithium abundance, chromospheric and coronal activity,
and fast rotation \citep{Gaidos98.1}.
                                                                                    
Two cool companions were discovered by \citet{Potter02.1}. The two objects, HD\,130948\,B and~C 
most likely form a gravitationally bound binary, and they both have spectral type L4 
\citep{Potter02.1, Goto02.1}.
HD\,130948\,B and C are unresolved with {\em Chandra} (separation $0.13^{\prime\prime}$),
and separated from the primary by only $2.6^{\prime\prime}$. 
Fig.~\ref{fig:acis_images} shows that HD\,130948\,BC is situated in the outer wing of the PSF of
the primary. There is no obvious enhancement of the count rate at the position of HD\,130948\,BC
with respect to other locations with the same distance to HD\,130948\,A. To examine the area
in more detail we have deconvolved the ACIS image with the PSF using the IDL
task {\sc max\_likelihood}\footnote{{\sc max\_likelihood.pro} is a tool provided by the 
IDL Astronomy User's Library, see http://idlastro.gsfc.nasa.gov/homepage.html}.  
No additional source is seen 
on the resulting reconstructed image 
at the position of HD\,130948\,BC,
and we conclude that the BD pair is not an X-ray emitter down to the sensitivity limit of this 
{\em Chandra} observation. We derive an upper limit of $\log{L_{\rm x}} < 25.6$\,erg/s. In
terms of \lglxlbol~ the upper limit is $-4.1$/$-4.2$, depending on whether it 
is attributed entirely to HD\,130948\,B or~C. 

The primary HD\,130948\,A is a bright X-ray source with roughly $0.2$\,cts/s in the $0.5-8$\,keV band, 
corresponding to a pile-up fraction of $\sim 20$\,\% according to Fig.~6.18 of the 
{\em Chandra} Proposers' Observatory Guide\footnote{http://cxc.harvard.edu/proposer/POG/index.html}. 
The negative effects of pile-up (distortion of the shape of the lightcurve and spectrum and underestimated 
flux) can be avoided by using an annular photon extraction region avoiding the core of the PSF,
obviously at the expense of statistics.  

\begin{figure}[t]
\begin{center}
\parbox{10.1cm}{
\parbox{6.5cm}{\resizebox{6.5cm}{!}{\includegraphics{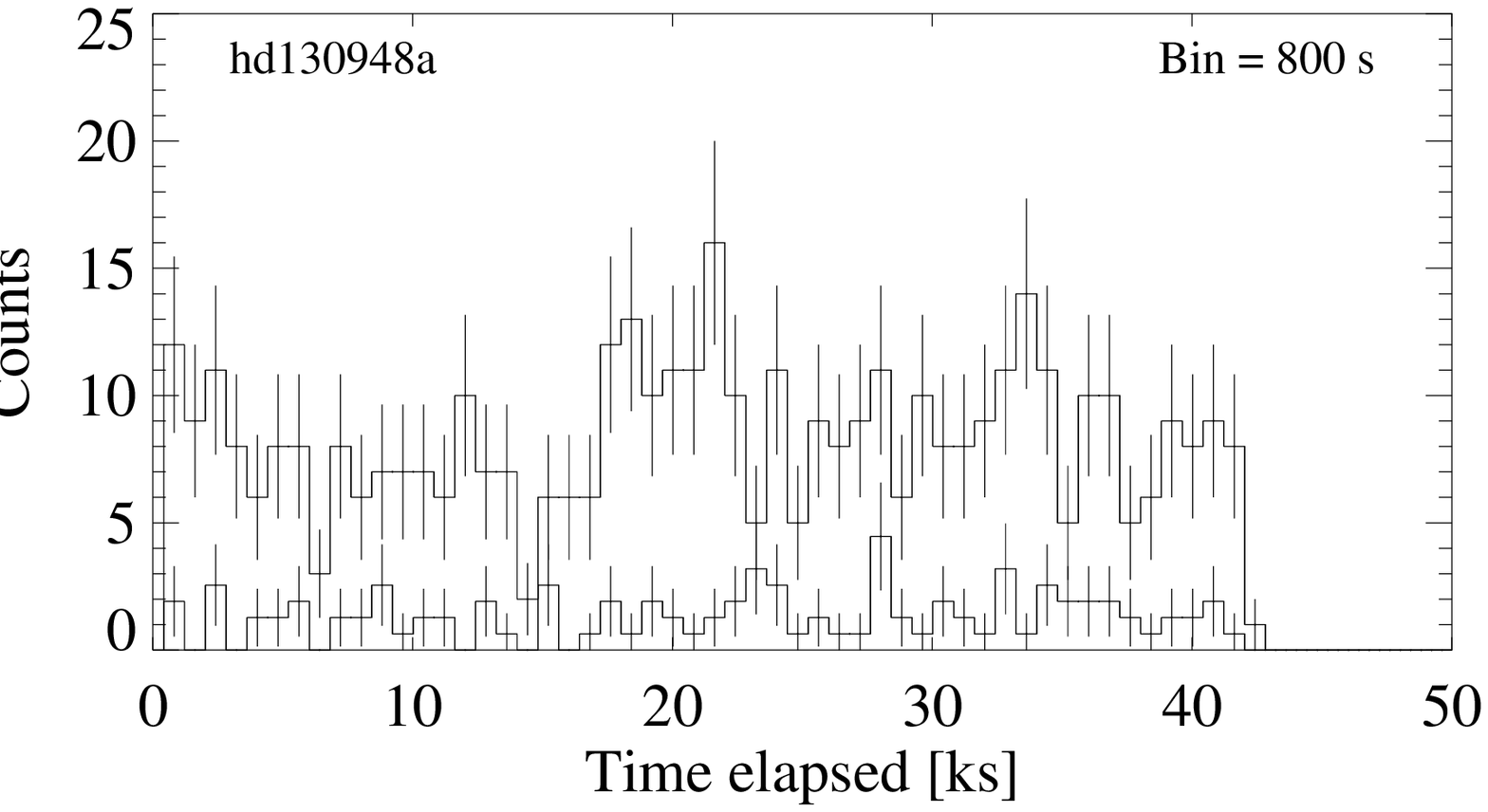}}}
\parbox{0.2cm}{ }
\parbox{2.6cm}{\resizebox{2.6cm}{!}{\includegraphics{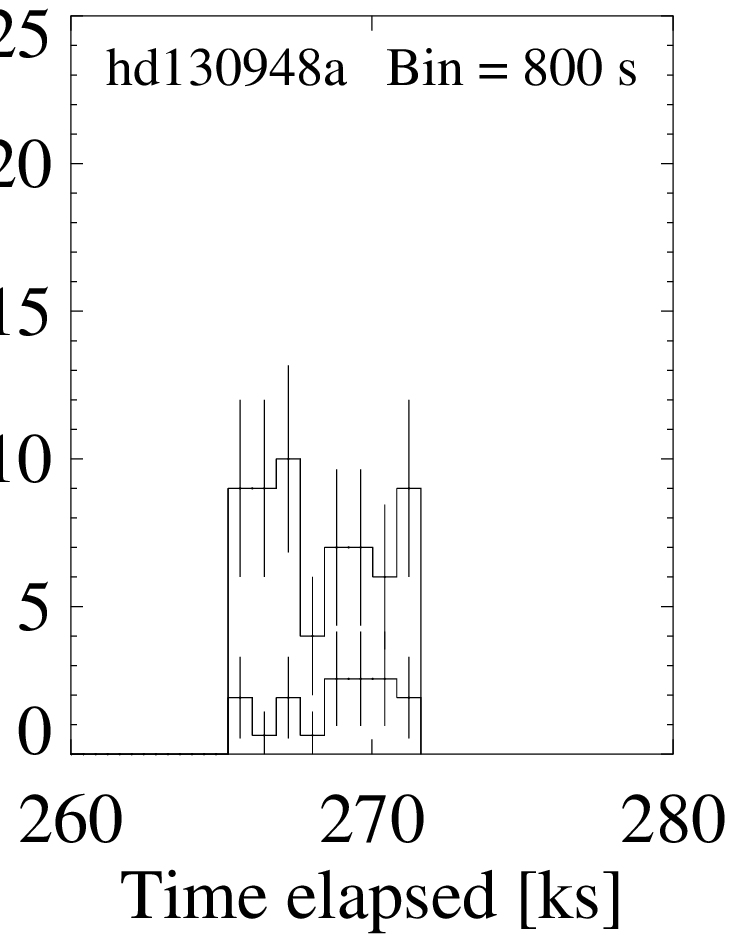}}}
}
\caption{X-ray lightcurve of HD\,130948\,A and background; source photons were extracted from an annulus with inner radius
of $1.5^{\prime\prime}$, thus avoiding the central portion of the PSF which is subject to pile-up.}
\label{fig:lc_hd130948}
\end{center}
\end{figure}

For the spectral analysis of HD\,130948\,A we consider only the longer observation, Obs-ID\,4471.
To determine the smallest inner radius $r_{\rm in}$ for an annular source area 
that reduces pile-up without losing too much statistics 
we extracted events from a series of annuli with $r_{\rm in} = 0.5^{\prime\prime}....3.75^{\prime\prime}$ 
and outer radius fixed at $8^{\prime\prime}$. As discussed above no significant contribution is present from the
BD pair, and there are also no other X-ray sources in the extraction region of HD\,130948\,BC.  
With help of the ChaRT\footnote{Information on the usage of the Chandra Ray Tracer (ChaRT) can be found at 
http://cxc.harvard.edu/chart/}
and MARX\footnote{The Model of AXAF Response to X-rays (MARX) is provided by the MIT/CXC 
(see http://space.mit.edu/CXC/MARX/)} simulators an individual {\em arf} was constructed for each of
the annuli to correct the effective area for the missing part of the
PSF\footnote{see http://www.astro.psu.edu/users/tsujimot/arfcorr.html and \citet{Getman05.1} 
for more information on the procedure.}.
The X-ray luminosity computed from the individual spectra increases as a function of $r_{\rm in}$ until 
it reaches a plateau when all of the piled-up portion of the PSF is excluded. The kink in the
distribution $L_{\rm x} - r_{\rm in}$ defines the optimized photon extraction radius.
From this plateau we determine a luminosity of $\log{L_{\rm x}} = 28.79 \pm 0.05$\,erg/s.
This is in reasonable agreement with the 
measurement by \citet{Huensch99.1} of $\log{L_{\rm x}} = 29.0$\,erg/s based on the {\em ROSAT}
All-Sky Survey, considering the known variability of $L_{\rm x}$ in solar-like stars
and the different characteristics of the satellites. 
The X-ray luminosity of HD\,130948\,A is also consistent with expectations for main-sequence
stars of similar rotation
periods \citep[$P_{\rm rot} = 7.85$\,d; ][]{Gaidos00.1} and age; see e.g. Fig.~3 in \citet{Pizzolato03.1}. 
The spectrum extracted from the optimized region can be described
with an iso-thermal APEC model of $kT \sim 0.5$\,keV ($\log{T}\,{\rm [K]} = 6.76$ = 5.8\,MK)
and subsolar global abundances of $Z = 0.5\,Z_\odot$. 
The {\em Chandra} lightcurve of HD\,130948\,A shows variability on timescales of
hours, but no strong flaring (Fig.~\ref{fig:lc_hd130948}). 
%
% OUTPUT OF    lc.pro     (prepare_lcplot.pro)
%

\subsection{Gl\,569}\label{subsect:gl569}

This is the only of the three targets where both the primary and the BD companion(s) are detected
(see Fig.~\ref{fig:acis_images}). The data analysis and the results for Gl\,569\,Bab have been 
described in detail by \citet{Stelzer04.2}. We summarize the results for the BD pair in Table~\ref{tab:sample}. 

Gl\,569\,A suffers from pile-up. We analysed its spectrum in an analogous way to that of HD\,130948\,A. 
The large flare that occurred $\sim 5$\,ksec after the start of the observation 
\citep[see Fig.~2 in ][]{Stelzer04.2} 
is eliminated from the quiescent spectrum and its spectrum is analysed separately. 
The mean X-ray luminosity (averaged over the full observation) is in accordance with
the {\em ROSAT} result \citep{Huensch99.1}. During the flare the temperature increased slightly from
$kT_{\rm Q} = 0.81$\,keV (lgT = 6.97; 9.4 MK) to $kT_{\rm F} = 0.90$\,keV (lgT = 7.02, 10.4 MK), 
as expected if additional heating is taking place during the outburst.

\section{Discussion}\label{sect:discussion}

\subsection{X-ray properties of brown dwarfs}\label{subsect:disc_xrayprop_bds}

For the two detected BDs from our sample, HR\,7329\,B and Gl\,569\,Bab, the X-ray spectra 
are rather similar. 
This is remarkable because HR\,7329\,B is a relatively young BD with low-amplitude variability, 
while Gl\,569\,Bab is relatively old and has shown a strong flare during the {\em Chandra} observation. 

Detailed characterizations of the X-ray spectra of BDs have remained out of reach to date due to
poor statistics: all observations undertaken so far have yielded a few dozen counts per source 
at most. Therefore, when comparing our sample to previous observations of BDs, 
we consider the median photon energy $\langle E \rangle$ 
as representative for the temperature of the emitting plasma. 
Comparing the time-averaged $\langle E \rangle$ we notice again the similarity of HR\,7329\,B and 
Gl\,569\,Bab. When computing $\langle E \rangle$ separately for the
flare and quiescent times of Gl\,569\,Bab it is seen that (i) during quiescence Gl\,569\,Bab is
softer than HR\,7329\,B, and (ii) the time-average of Gl\,569\,Bab is dominated by the flare. 
The values of $\langle E \rangle$ for our BD targets are somewhat lower
than for the younger BDs in IC\,348 \citep[see Fig.~5 of][]{Preibisch02.1}, suggesting a softer X-ray spectrum. 
However, one must keep in mind that the X-ray spectrum of young BDs suffers from extinction. 
The X-ray detected BDs in IC\,348 have a range of $A_{\rm J} = 0.2 ... 1.8$\,mag, with the majority of
them at $A_{\rm J} \sim 0.4$\,mag, corresponding to 
$A_{\rm V} \sim 1.4$\,mag and $N_{\rm H} \sim 0.3\,10^{22}\,{\rm cm^{-2}}$. 
Similarly, \citet{Preibisch05.1} found $\langle E \rangle \sim 1...2$\,keV for BDs in the ONC. 
These BDs are absorbed by $A_{\rm V} = 1.4 ....6.8$\,mag. 

\begin{figure}[t]
\begin{center}
\resizebox{9cm}{!}{\includegraphics{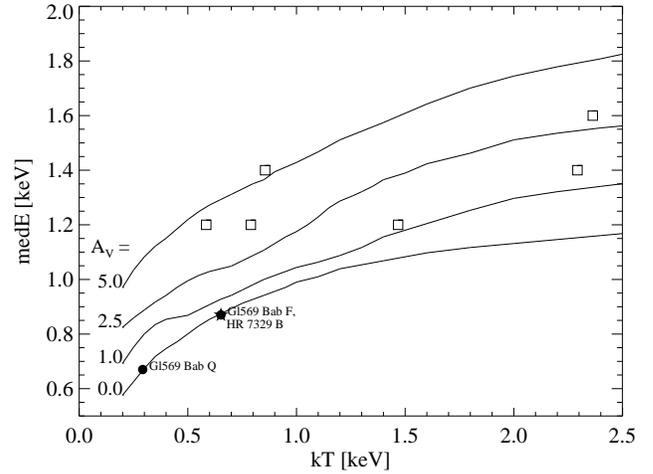}}
\caption{Median photon energy vs. X-ray temperature obtained from simulations of a 1-T APEC spectrum with
photo-absorption. The individual curves correspond to spectra with different column density $N_{\rm H}$.
We assumed a standard extinction law of $N_H$[cm$^{-2}] = 2.0~10^{21}$ A$_V$[mag] \citep{Ryter96.1}. Squares
denote BDs in the ONC from \protect\citet{Preibisch05.1}.}
\label{fig:medE}
\end{center}
\end{figure}

We have simulated spectra with XSPEC on a grid of $kT$ and $N_{\rm H}$ values, and computed
$\langle E \rangle$ for each of the theoretical spectra. In Fig.~\ref{fig:medE} we show the 
resulting relation between the median photon energy and the temperature of the 1-T model for
different values of $N_{\rm H}$. 
If for a given BD the absorption is known the X-ray temperature can be estimated from $\langle E \rangle$. 
We use the only two BDs for which both $\langle E \rangle$ and $kT$ have been measured,
Gl\,569\,Bab and HR\,7329\,B, as a cross-check of this procedure, and, 
indeed, we obtain $kT$ compatible with the results from spectral fitting of the data; 
compare Fig.~\ref{fig:medE} to text in Sect.~\ref{sect:results}. 
We estimate in the same way $kT$ for the BDs in the ONC using extinction and median photon energies
given by \citet{Preibisch05.1}. Fig.~\ref{fig:medE} shows that for these objects the rather 
small range of measured $\langle E \rangle$
corresponds to a large spread in $kT$, because of their wide range of absorption.
Some of the very young BDs in the ONC apparently have soft spectra like the more evolved BDs we discuss
in this paper. This is unlike in stars, where X-ray temperature decreases with increasing age 
\citep{Favata03.1}.
However, some other ONC BDs are much harder. 
Unfortunately, the ONC BD sample is to our knowledge the only one for which the parameter 
$\langle E \rangle$ was presented, and therefore at this point 
we can not draw firm conclusions on the evolution of coronal temperature with age in BDs. 
%
% OUTPUT FROM    plot_lx_rin.pro
%

Our third target, HD\,130948\,BC, is undetected with an upper limit to $L_{\rm x}$ three orders of magnitude
lower than the luminosity determined for its primary HD\,130948\,A. 
We mention in passing that \cite{Tsuboi03.1} have studied a similar case: TWA-5\,B, which was detected
and resolved with
{\em Chandra} from its primary. The two components in that system have an even smaller separation of
$2^{\prime\prime}$. But TWA-5\,A is `only' two orders of magnitude brighter in X-rays than the BD, 
and TWA-5\,B has an at least $\sim 10$ times larger flux than HD\,130948\,B.

\subsection{BD X-ray emission and stellar parameters}\label{subsect:disc_xraystel_bds}

The new {\em Chandra} observations 
put us for the first time in the position to study the X-ray properties 
of BDs in restricted ranges of mass and temperature. The parameter space that can be  
studied on basis of the existing data is high-lighted by grey shades in Fig.~\ref{fig:teff_age}:   
the mass range $0.05-0.07\,M_\odot$ and the $T_{\rm eff}$ range $2400-2800$\,K. 

When comparing X-ray luminosities for BDs obtained from different observations, we must
take into account that different authors have used different energy bands for the evalution
of $L_{\rm x}$. We have used PIMMS to convert all $L_{\rm x}$ values from the literature 
to the $0.5-8$\,keV band. The assumed spectral form is a Raymond-Smith model with individual
temperature based on the results presented in the literature with the exception of DENIS\,J1556.
\citet{Bouy04.1} has assumed an X-ray temperature of $2$\,keV for this BD in Upper Sco, but 
we suspect that the actual temperature might be substantially lower, such as observed on 
TWA-5\,B and on HR\,7329\,B which are only slightly older than DENIS\,J1556. If the temperature 
of DENIS\,J1556 is $0.4$\,keV the X-ray luminosity decreases by a factor of $0.7$ with respect
to the value for $2$\,keV. 
In this study, instead of the published value, 
we use $\log{L_{\rm x}} = 28.1$\,erg/s for DENIS\,J1556, derived from a recent (re-)analysis of the 
same {\em XMM-Newton} data (Argiroffi et al., in prep.). This is smaller by a factor $4$
than the estimate by \citet{Bouy04.1}, and yields a \lglxlbol~ ratio of $-3.3$, more typical
for coronal emission than the published value of $-2.69$.

\subsubsection{High-mass brown dwarfs ($0.05-0.07\,M_\odot$)}\label{subsubsect:hm}

\begin{figure*}
%\sidecaption 
\includegraphics[width=16cm]{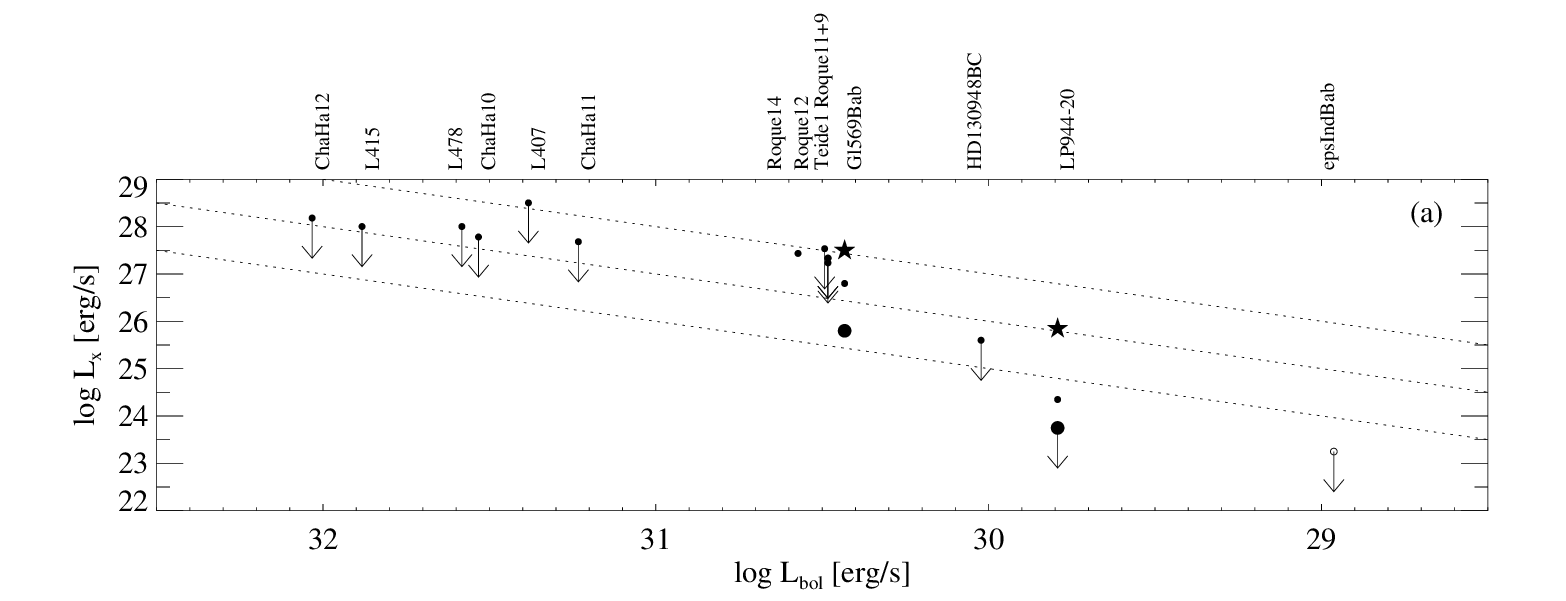}
\includegraphics[width=16cm]{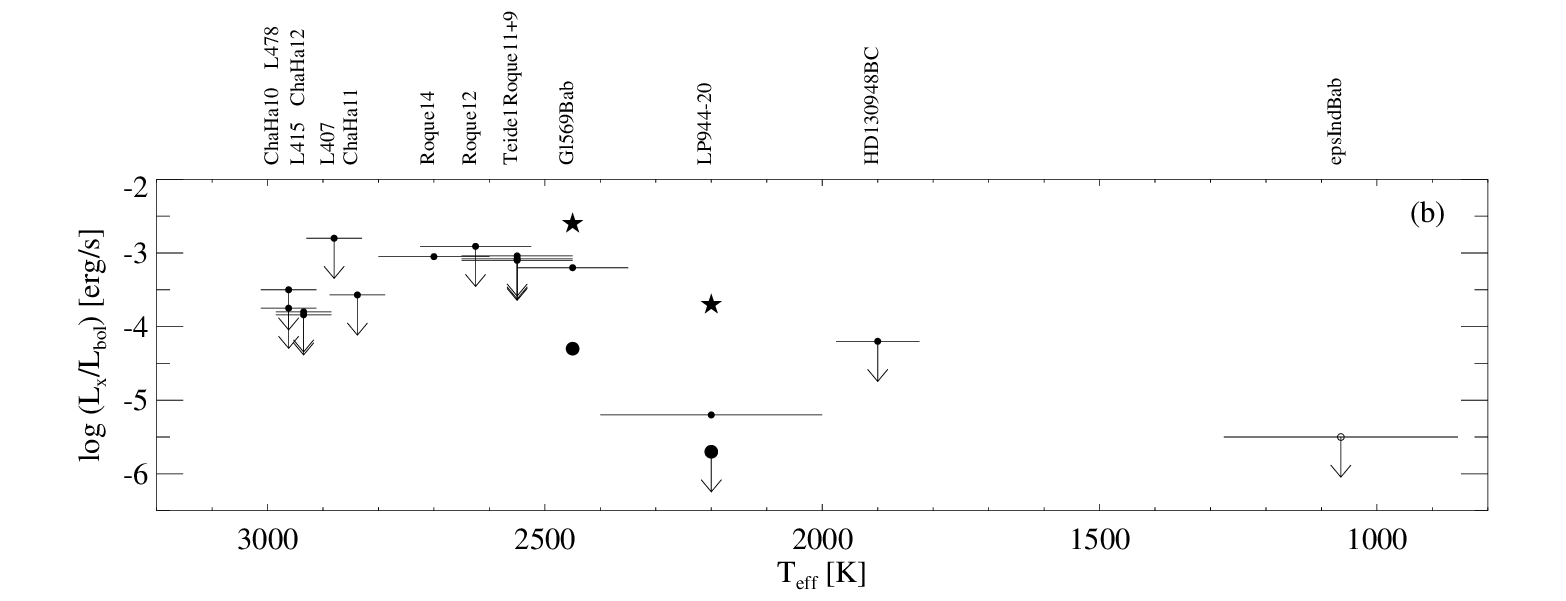}
\includegraphics[width=16cm]{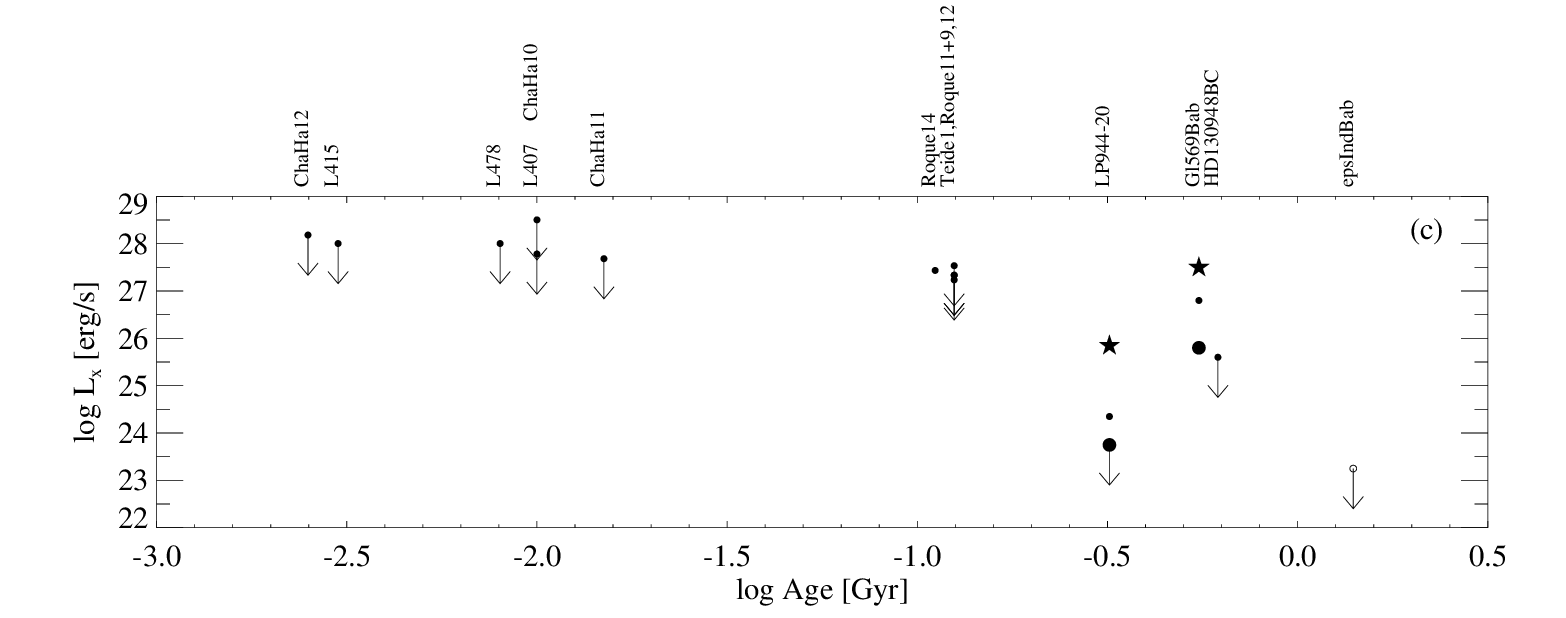}
\caption{X-ray activity of BDs with $M \sim 0.05 - 0.07\,M_\odot$ 
versus $L_{\rm bol}$ (a), versus $T_{\rm eff}$ (b) and versus age (c). 
Plotting symbols: 
{\em small circles} - Time-averaged $L_{\rm x}$ from {\em Chandra} and {\em XMM-Newton} observations; 
{\em star symbols} - $L_{\rm x}$ in the flare peak; 
{\em large circles} - Quiescent $L_{\rm x}$. Dotted lines in the uppermost panel 
indicate $L_{\rm x}/L_{\rm bol}$ levels of $10^{-3}$, $10^{-4}$, and $10^{-5}$.} 
\label{fig:xrayevol_m06}
\end{figure*}

First we examine how the X-ray emission of BDs with the same mass ($0.05-0.07\,M_\odot$; 
shaded dark grey in Fig.~\ref{fig:teff_age}) is related to bolometric luminosity and to
effective temperature.
We have included $\epsilon$\,Ind\,Bab in this `high-mass' BD sample 
because it is the only object with sensitive X-ray data
that allows us to extend the temperature range significantly below $2000$\,K. But let's keep in mind 
that the mass of both components in this BD binary are slightly lower than the lower threshold
of the studied mass bin 
($M_{\rm \epsilon Ind Ba} \sim 0.045\,M_\odot$; $M_{\rm \epsilon Ind Bb} \sim 0.025\,M_\odot$; 
McCaughrean et al. 2004). 
Therefore, $\epsilon$\,Ind\,Bab is shown with an open plotting symbol
in Fig.~\ref{fig:xrayevol_m06}. 
 
Fig.~\ref{fig:xrayevol_m06}\,a shows the relation between $L_{\rm x}$ and $L_{\rm bol}$
for the sample of $0.05-0.07\,M_\odot$ objects. 
For unresolved binaries we have summed up the bolometric luminosities of both components. 
Most data points in Fig.~\ref{fig:xrayevol_m06}\,a fall into the $10^{-3...-5}$ range
typical for late-type active stars. But the slope seems to steepen for the least luminous
objects which correspond to older ages. In other words, there is a decline of the 
$L_{\rm x}/L_{\rm bol}$ ratio as the BDs evolve; see Fig~\ref{fig:xrayevol_m06}\,c
discussed in more detail below.   

Fig.~\ref{fig:xrayevol_m06}b shows the $L_{\rm x}/L_{\rm bol}$ ratio as a function 
of $T_{\rm eff}$ for the same sample. 
Despite the large number of upper limits, there is evidence for a decline of the 
fractional X-ray luminosity ($L_{\rm x}/L_{\rm bol}$) with $T_{\rm eff}$. 
This underlines that the atmospheric temperature plays a crucial role in determining 
the level of X-ray activity of evolved BDs. 
Currently, there are only two data points below $T_{\rm eff} \sim 2000$\,K, i.e. in the regime
of the L and T dwarfs, and both are non-detections. 
Given that the data set is dominated by upper limits, we have refrained from a more 
detailed analysis involving statistical tests. 

From Figs.~\ref{fig:xrayevol_m06}a and~\ref{fig:xrayevol_m06}b it strikes that 
none of the IC\,348 and Cha\,I BDs in the $0.05-0.07\,M_\odot$ mass bin was detected in X-rays.
As can be seen from Fig.~\ref{fig:teff_age} all but one of the X-ray detected objects 
from the IC\,348 and Cha\,I sample
have either masses above the substellar limit or they are extremely young 
(above the model grids), when compared to \cite{Baraffe98.1} and \cite{Chabrier00.1} models.
On the other hand, four of the six IC\,348 and Cha\,I members in our $0.05-0.07\,M_\odot$ bin 
have ages of $\sim 10$\,Myr. This can be seen from Fig.~\ref{fig:xrayevol_m06}\,c, which shows
the evolution of X-ray luminosity with age. 

The observed $L_{\rm x}/L_{\rm bol}$ values of detected BDs at similar $T_{\rm eff}$ but lower
or higher mass than the range examined in Fig.~\ref{fig:xrayevol_m06} 
are comparable to the observed upper limits, such that the emission levels of
the undetected young BDs in the $0.05-0.07\,M_\odot$ bin are expected to be only 
little below their upper limits. 
Therefore, even a wrong assignment of masses for the young
BDs by the evolutionary models -- e.g. the \citet{DAntona97.1} models assign systematically lower
masses for BDs of given $T_{\rm eff}$ than the \citet{Baraffe98.1} and \citet{Chabrier00.1} models 
-- does not change our consideration. 

%
% OUTPUT FROM     plot_xrayevol.pro
%

\subsubsection{Hot Brown dwarfs ($T_{\rm eff} = 2400 - 2800$\,K)}\label{subsubsect:hot}

\begin{figure*}
%\sidecaption
\includegraphics[width=16cm]{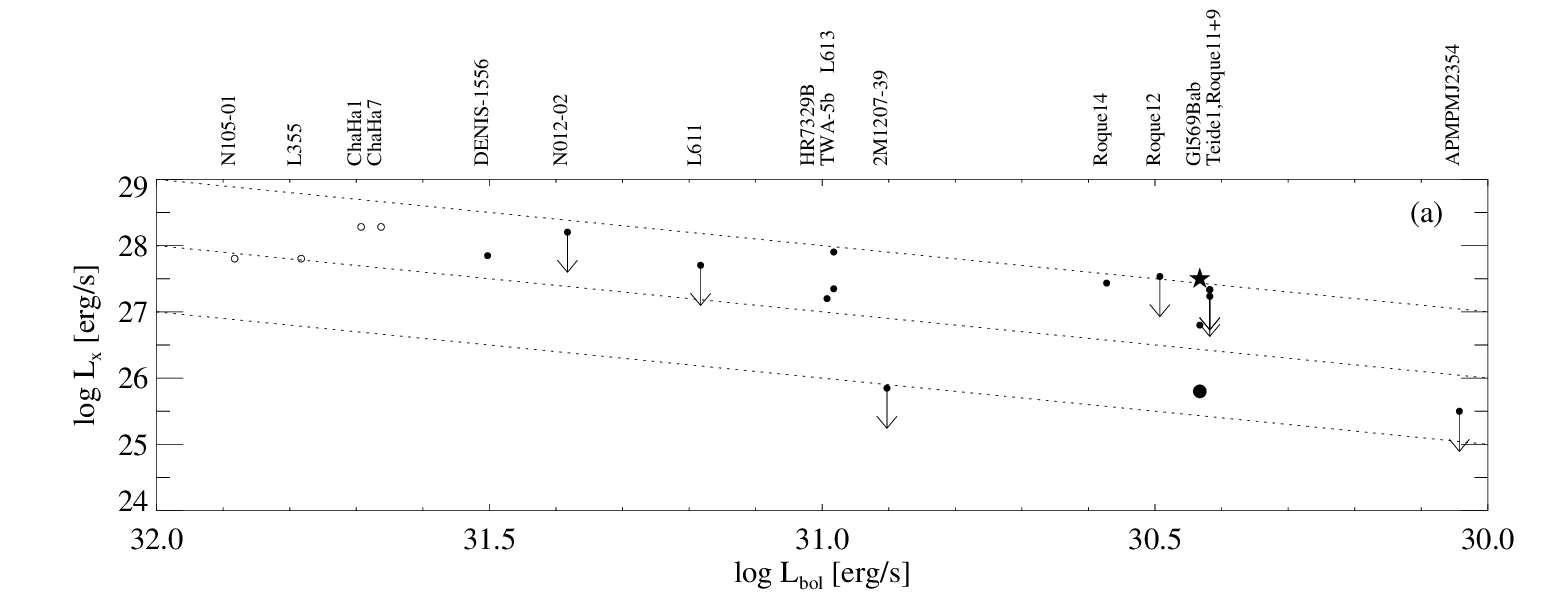}
\includegraphics[width=16cm]{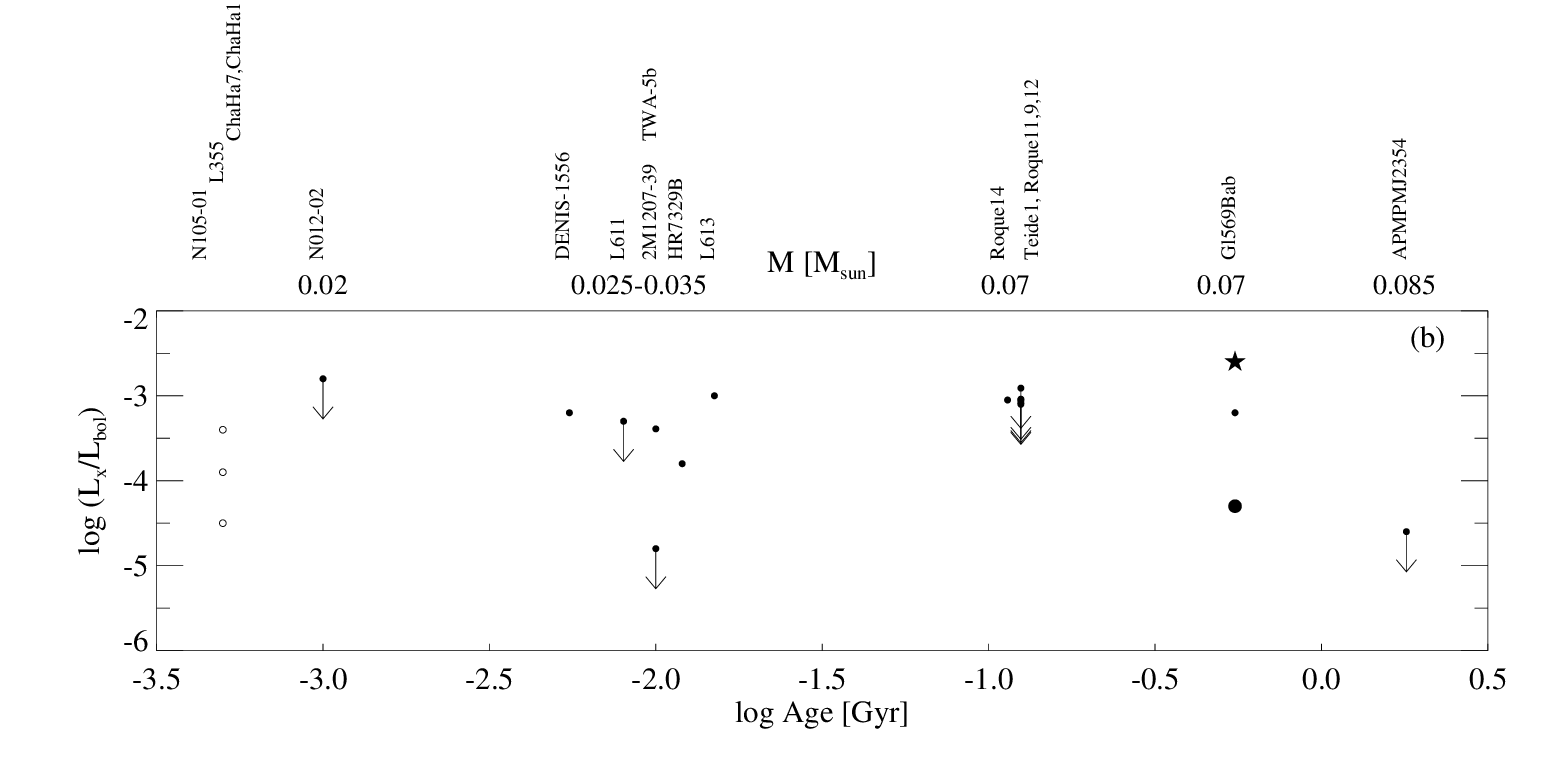}
\caption{X-ray activity for BDs with $T_{\rm eff} = 2400...2800$\,K: 
top -- $L_{\rm x}$ versus $L_{\rm bol}$, bottom -- $L_{\rm x}/L_{\rm bol}$ ratio versus age.
BDs above the \citet{Baraffe98.1} and \citet{Chabrier00.1} models in the HR diagram are 
arbitrarily placed at an age of $0.5$\,Myr, and given open plotting symbols.} 
\label{fig:xrayevol_t2600}
\end{figure*}

The dependence of X-ray luminosity on $L_{\rm bol}$ for BDs in a given range of $T_{\rm eff}$ 
($2400-2800$\,K; see light grey shade in Fig.~\ref{fig:teff_age}) 
is displayed in Fig.~\ref{fig:xrayevol_t2600}\,a. 
Most BDs fall in a narrow band defined by $L_{\rm x}/L_{\rm bol} = 10^{-3...-4}$,
indicating that the X-ray activity levels for `hot' BDs are 
(A) similar irrespective of age and mass, and (B) near the upper end of the canonical 
values for late-type stars. 

The dependence of X-ray activity on age for the same $T_{\rm eff}$ bin is examined in 
Fig.~\ref{fig:xrayevol_t2600}\,b.  
Here, the most notable feature is the 
low upper limit of 2M\,1207-39, which deviates by nearly 
two orders of magnitude from the other BDs at the same age of $\sim 10$\,Myr. 
Note, that all objects in this age range 
have similar mass of $\sim 0.025-0.035\,M_\odot$. 
Accretion may be an additional parameter that influences the observed X-ray emission in 
young BDs. Only two BDs among the objects shown in Fig.~\ref{fig:xrayevol_t2600} display
signatures for accretion, 2M\,1207-39 and L355;  
see discussion in Sect.~\ref{subsect:disc_acc}. 
%
% OUTPUT FROM    plot_xrayevol.pro
%

\subsection{BD X-ray emission and accretion}\label{subsect:disc_acc}

%
% OUTPUT FROM     plot_lx_lbol_bds.pro
%
\begin{figure}
\begin{center}
\resizebox{8cm}{!}{\includegraphics{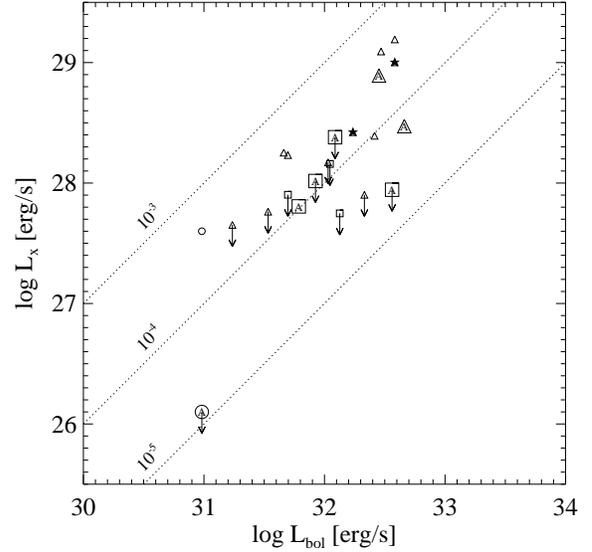}}
\caption{
Ratio of $\log{L_{\rm x}}$ versus $\log{L_{\rm bol}}$ for young BDs and BD candidates
(spectral type M5 and later): 
{\em Triangles, squares, open circles} -- BDs in Cha\,I, IC\,348, TW\,Hya. 
The plot includes objects with spectral type M5 and later with published measurement of
H$\alpha_{\rm 10}$. Accretors are marked with `A' and a larger plotting symbol. Two objects
in Cha\,I that have flared during the X-ray observation are marked with an asterisk.}
\label{fig:lxlbol}
\end{center}
\end{figure}

In Fig.~\ref{fig:lxlbol} we show only the young BDs (IC\,348, Cha\,I, TW\,Hya)   
in the $L_{\rm x} - L_{\rm bol}$ diagram with the purpose 
of examining the influence of accretion on X-ray emission from BDs. 
We distinguish accreting from non-accreting objects by a full width at $10$\,\% of the peak flux of 
H$\alpha$ (H$\alpha_{\rm 10}$) larger than $200$\,\AA; following \citet{Jayawardhana03.2}. 
This requires that the sample is restricted to BDs with measurements of H$\alpha_{\rm 10}$. 
The H$\alpha$ data is taken 
from the literature (\cite{Jayawardhana03.2}, \cite{Natta04.1}, 
\cite{Muzerolle05.1} and \cite{Mohanty05.1}  
for the star forming regions, and \cite{Mohanty03.2} for TWA).  
It turns out that H$\alpha_{\rm 10}$ is available for only $7$ out of
$19$ young bona-fide BDs, and 
only two of them (L\,355 in IC\,348 and 2M\,1207-39 in TWA) %, ChaH$\alpha$\,11 in Cha\,I,
are identified as accreting. 

A broader statistical basis is obtained when including BD candidates, i.e. objects with
spectral type M5 to M7. 
In this sample $22$ objects have both sensitive X-ray measurements and H$\alpha_{\rm 10}$ data,
and $7$ are above the $200$\,\AA~ threshold for accretion. 
The accretors are identified by an `A' surrounded by a larger symbol in Fig.~\ref{fig:lxlbol}. 
There may be some trend for the accretors to show lower $L_{\rm x}/L_{\rm bol}$ ratio than
the non-accreting young BDs, but admittedly this assertion needs to be corroborated by
a much larger sample. We deem worth noting, however, that the two BDs in TW\,Hya 
behave very differently both in X-rays and in H$\alpha$:
TWA-5\,B is a rather bright X-ray source with moderate H$\alpha$ emission 
\citep[$W_{\rm H\alpha} = 5$\,\AA, H$\alpha_{\rm 10} = 162$\,\AA;][]{Mohanty03.2},
while 2M\,1207-39 is undetected in X-rays down to a rather sensitive upper limit 
\citep{Gizis04.1}, but
is known to exhibit broad and strong H$\alpha$ emission 
\citep[$W_{\rm H\alpha} = 28$\,\AA, H$\alpha_{\rm 10} = 204$\,\AA;][]{Mohanty03.2}. 

The H$\alpha$ profile of 2M\,1207-39 was recently shown to be strongly variable 
with $10$\,\% width ranging from $200...300$\,\AA \citep{Scholz05.1}. 
This leads us to caution that the number of accretors in our
sample might actually be a lower limit, and one or more of the BDs below the $200$\,\AA~ threshold
may have been caught in a temporary low-accretion phase during the H$\alpha$ observations. 
Furthermore, for an ideal comparison X-ray and optical data should be obtained simultaneously. 
However, the X-ray
emission of young BDs (and also that of higher-mass T Tauri stars), undergoes only modest changes on a
long time baseline \citep[e.g.][]{Stelzer04.1}. Therefore, we consider the present data set as
representative for young BDs and 
speculate that
accretion may suppress X-ray emission in BDs, similar to what is observed in higher-mass T Tauri 
stars \citep[see e.g.][]{Stelzer01.1, Flaccomio03.1}. 
Several mechanisms by which accretion may cause a decrease of X-ray emission in T Tauri stars have been
suggested, including distortion of the magnetosphere, increased coronal density, and
changes in the interior structure that influence the dynamo process \citep[see][]{Preibisch05.2}.

\subsection{X-ray properties of the primaries}\label{subsect:disc_xrayprop_prim}

\begin{figure}[t]
\begin{center}
\resizebox{9cm}{!}{\includegraphics{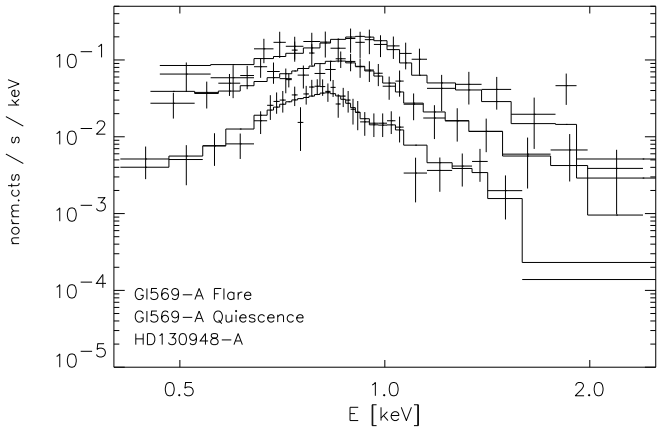}}
\caption{Spectra of the primaries, from top to bottom Gl\,569\,A during flare phase, 
Gl\,569\,A during quiescence, and average of HD\,130948\,A. The spectra of Gl\,569\,A
are offset in the vertical direction for clarity. The enhanced bin at $\sim 1.85$\,keV 
in the flare spectrum of Gl\,569\,A may point at strong Si\,XIII emission.} 
\label{fig:spec_prims}
\end{center}
\end{figure}

The nature of the primary was not a selection criterion for the targets of this
study, except for their age. Therefore it is not a surprise that they are very 
different from each other. 
It is nevertheless interesting to examine and compare their X-ray properties. 
The ACIS spectra of all detected primaries of our sample are shown in 
Fig.~\ref{fig:spec_prims}, and the derived X-ray parameters are summarized in 
Table~\ref{tab:sample_prim}. As a consequence of the special photon extraction procedure
required to avoid pile-up, each of the spectra of the detected primaries 
has only $\sim 500$\,counts, such that a detailed analysis of their temperature 
and emission measure distribution is not feasible. 

Nevertheless, the temperatures from the iso-thermal fits can be considered as an 
average coronal temperature,
and be compared to the recent work by \citet{Telleschi05.1}. 
Their study of solar-analogs is based on high-resolution X-ray spectra,
and they defined a mean coronal temperature 
as the average of the temperatures from the emission measure distribution weighted by the 
emission measure. 
In order to compare these different data, 
we must take into account that different energy bands were used to measure
the X-ray luminosities: 
$0.5-8$\,keV in our case versus $0.1-10$\,keV by \citet{Telleschi05.1}. 
We verified with PIMMS that this amounts to a flux change by a factor of two 
for a $1$\,keV source. After adapting the relation between $T_{\rm x}$ and $L_{\rm x}$ 
observed by \citet{Telleschi05.1} for a sample of $6$ solar-analogs to the
$0.5-8.0$\,keV band, the predicted X-ray luminosity for HD\,130948\,A is
$10^{29.0}$\,erg/s, versus the observed value of $10^{28.8}$\,erg/s. Therefore, the G2 star HD\,130948\,A 
behaves like a typical solar-analog, 
providing further support for the tight relation between $L_{\rm x}$ and $T_{\rm x}$ in similar stars.
Gl\,569\,A, on the other hand, is a much cooler, lower-mass flare star 
(spectral type dM2e), and, as expected, 
clearly does not conform to the correlation observed for the G stars. 
%
% OUTPUT FROM        /home/stelzer/data/IDL/allg/plot_acis.pro   (prepare_specplot.pro)
%

The non-detection of HR\,7329\,A is 
in agreement with the common understanding that intermediate-mass stars can 
sustain neither coronal nor wind-driven X-ray emission. Contrary to this 
expectation, many A-type stars have been detected in X-ray observations with various
satellites. However, most of these observations were carried out with instruments with low spatial
resolution and cannot exclude the likelihood that the observed emission is to be attributed to nearby
unresolved companions (as proven here in the case of HR\,7329), or unrelated nearby objects. 
The very low upper limit we establish for HR\,7329\,A is a strong constraint for any
activity resulting from eventual core dynamo action \citep{MacGregor03.1}.

\section{Conclusions}\label{sect:conclusions}

Our {\em Chandra} targets, combined with data from the literature, represent a so far unique sample that 
has allowed us to make a first quantitative study of 
the dependence of sub-stellar X-ray emission on critical parameters.

From Fig.~\ref{fig:xrayevol_t2600} it appears to emerge that 
BDs in young associations ($\beta$\,Pic; TWA), open clusters (Pleiades) and the
solar neighborhood (Gl\,569\,Bab) that share the same temperature ($2400-2800$\,K) 
have very similar fractional X-ray luminosities 
($\log{(L_{\rm x}/L_{\rm bol})} \sim 10^{-3...-4}$), i.e. the X-ray luminosity
scales with the bolometric luminosity for `hot' BDs.

Note, that for the older objects this activity level corresponds to higher mass. 
This phenomenon can probably be extended to the coolest stars above the substellar boundary: 
MS stars with $T_{\rm eff}$ in the range examined in Fig.~\ref{fig:xrayevol_t2600},
i.e. stars with spectral types later than $\sim$ M7, 
show $\log{(L_{\rm x}/L_{\rm bol})}$ levels similar to our `hot' BD sample  
\citep[see e.g. Fig.4 of][]{Stelzer04.2}, and -- in general -- such stars are thought to 
have mature ages of a few Gyr. 
But, strictly speaking, they can not be included in our study because actual 
age determinations are not available.  

However, in the sample restricted to high-mass BDs 
($0.05-0.07\,M_\odot$; the only substellar mass range covered so far with sensitive
X-ray observations) $L_{\rm x}$ seems to decrease with $L_{\rm bol}$ steeper than 
expected for the canonical \lglxlbol~ ratio of late-type stars ($\sim 10^{-3...-5}$). 
This sample includes BDs with low atmospheric temperature.
And, indeed, an interesting fall of the $L_{\rm x}/L_{\rm bol}$ ratio 
with $T_{\rm eff}$ is observed (Fig.~\ref{fig:xrayevol_m06}\,b).

With the present data it is not possible to unambiguously distinguish the influence of
$L_{\rm bol}$ and $T_{\rm eff}$ on the X-ray activity level. But we have presented first 
evidence for a direct correlation between $L_{\rm x}/L_{\rm bol}$ and $T_{\rm eff}$. Such a correlation 
supports conclusions based on calculations of magnetic diffusivity. \citet{Mohanty02.1}  
found that the high electrical resistivities in the predominantly neutral atmospheres of
ultracool dwarfs prevent significant dynamo action by (a) reducing the amount of magnetic stress
created by footpoint motions and (b) impeding their efficient transport through the atmosphere. 
In this vein, improved statistics are required for the critical temperature range, i.e. the regime of 
late-M and L dwarfs.  

HD\,130948\,BC is the only X-ray observed L dwarf with known age, and probably the youngest one. 
Unfortunately, its upper limit is rather high due to 
contamination with photons of the nearby X-ray bright primary. 
Together with all other L dwarfs studied with reasonable sensitivity \citep{Stelzer02.1, Berger05.1} it is
undetected, such that no L dwarf has been detected in X-rays so far.
The situation is thus qualitatively similar to what is found for chromospheric H$\alpha$
activity: H$\alpha$ emission shows a sharp drop at the latest M types, albeit still a significant
fraction of early L dwarfs has H$\alpha$ in emission \citep{Gizis00.1, Mohanty03.1}. 
But this apparent discrepancy to the X-ray observations might be due to the
much larger data base for chromospheric emission in ultracool dwarfs as opposed to coronal
emission. 

We examined the incidence of accretors in the X-ray sample
of young BDs, and find marginal evidence that accreting BDs have lower X-ray luminosity than non-accreting 
ones. If confirmed by future studies on a larger sample this would show that activity on BDs 
is subject to the same mechanisms that suppress X-ray emission in 
pre-MS stars during their T Tauri phase.

With the presentation of two X-ray spectra for a sample of three evolved BDs observed with
{\em Chandra} we have demonstrated the high potential of this X-ray satellite for the study of
substellar coronae. 
Similar values for X-ray temperature are measured for the $12$\,Myr old HR\,7329\,B 
during a relatively quiescent
state and for the $\sim 200-400$\,Myr old Gl\,569\,Bab during a large flare. 
A systematic study of temperatures of the X-ray emitting plasma 
in BDs is impeded by the absence of similarly high-quality data in the previous X-ray literature. 
But we have shown that the median photon energy can be used to infer the X-ray temperature of
the corresponding spectrum, if the absorption is known. 
This way, even with low photon statistics the coronal plasma of BD samples 
representing different evolutionary phases can be compared.
From the median photon energy 
we deduce a temperature of Gl\,569\,Bab during quiescence of $\sim 0.3$\,keV, much
lower than during the flare, and much lower than the temperature of HR\,7329\,B during
quiescence, thus confirming the presumptions that (a) the coronal temperature decreases 
as BDs evolve, (b) this does not prevent the occurrence of large outbursts.  

Our study has shown that despite the increasing data base on X-ray emission
from BDs, the sensitivity of the existing observations is not always satisfactory. 
In future {\em Chandra} and {\em XMM-Newton} observations 
we hope to obtain better detection rates and to extend further 
the coverage of the age-mass-luminosity-temperature parameter space,
to examine the influence of rotation, and to constrain the coronal temperatures 
near and beyond the substellar regime.

\begin{acknowledgements}

We would like to thank the referee K. Briggs for inspiring us to  
a critical scrutiny of the results, and for his care in reading the paper. 
BS acknowledges financial support from MIUR PRIN 2004-2006. 
Support for this work was provided by the National Aeronautics and Space Administration 
through Chandra Award Number
05200031 issued by the Chandra X-ray Observatory Center, which is operated by the Smithsonian Astrophysical 
Observatory for and on behalf of the National Aeronautics Space Administration under contract NAS8-03060.

\end{acknowledgements}

\end{document}